\documentclass[conference,compsoc]{IEEEtran}
\ifCLASSOPTIONcompsoc
  \usepackage[nocompress]{cite}
\else
  \usepackage{cite}
\fi
\ifCLASSINFOpdf
\else
\fi
\usepackage{amsmath,amssymb,amsfonts}
\newtheorem{definition}{Definition}

\usepackage{bm}
\usepackage{cite}
\usepackage{bbding}
\usepackage{wasysym}
\usepackage{algorithm}
\usepackage[noend]{algpseudocode}
\usepackage{multirow}
\usepackage{mathrsfs}

\usepackage{tikz}
\usepackage{savesym}
\savesymbol{iint}
\savesymbol{iiint}
\usepackage{amsmath}

\usepackage[numbers, sort&compress]{natbib}

\usepackage{bbding}

\usepackage{filecontents}

\usepackage{bbding}
\usepackage{wasysym}
\usepackage{booktabs}
\usepackage{makecell}

\usepackage{bbm}

\newtheorem{theorem}{Theorem}

\usepackage{tabu}
\usepackage{threeparttable}
\usepackage{tabularx}

\usepackage{hyperref}
\usepackage{url}

\usepackage{color}
\newcommand{\mtrevise}{\textcolor{black}}
\newcommand{\mainmtrevise}{\textcolor{black}}

\usepackage{enumitem}

\usepackage{balance}

\begin{document}
%

\title{Securing Graph Neural Networks in MLaaS: \\A Comprehensive Realization of Query-based Integrity Verification}

\author{\IEEEauthorblockN{Bang Wu\IEEEauthorrefmark{1}\IEEEauthorrefmark{2}\textsuperscript{1},
Xingliang Yuan\IEEEauthorrefmark{2}, 
Shuo Wang\IEEEauthorrefmark{3},
Qi Li\IEEEauthorrefmark{4},
Minhui Xue\IEEEauthorrefmark{1} and
Shirui Pan\IEEEauthorrefmark{5}
}
\IEEEauthorblockA{\IEEEauthorrefmark{1}CSIRO's Data61, Australia}
\IEEEauthorblockA{\IEEEauthorrefmark{2}Monash University, Australia}
\IEEEauthorblockA{\IEEEauthorrefmark{3}Shanghai Jiao Tong University, China}
\IEEEauthorblockA{\IEEEauthorrefmark{4}Tsinghua University, China}
\IEEEauthorblockA{\IEEEauthorrefmark{5}Griffith University, Australia}}

\maketitle

\begin{abstract}
    The deployment of Graph Neural Networks (GNNs) within Machine Learning as a Service (MLaaS) has opened up new attack surfaces and an escalation in security concerns regarding model-centric attacks. 
    These attacks can directly manipulate the GNN model parameters during serving, causing incorrect predictions and posing substantial threats to essential GNN applications. 
    Traditional integrity verification methods falter in this context due to the limitations imposed by MLaaS and the distinct characteristics of GNN models.
    
    In this research, we introduce a groundbreaking approach to protect GNN models in MLaaS from model-centric attacks. 
    Our approach includes a comprehensive verification schema for GNN's integrity, taking into account both transductive and inductive GNNs, and accommodating varying pre-deployment knowledge of the models. 
    We propose a query-based verification technique, fortified with innovative node fingerprint generation algorithms. 
    To deal with advanced attackers who know our mechanisms in advance, we introduce randomized fingerprint nodes within our design. 
    The experimental evaluation demonstrates that our method can detect five representative adversarial model-centric attacks, displaying 2 to 4 times greater efficiency compared to baselines.
\end{abstract}
\IEEEpeerreviewmaketitle

\thispagestyle{plain}

\section{Introduction}
Graph Neural Networks (GNNs) have drawn considerable attention due to their demonstrated ability to analyze graph data~\cite{KipfW17,0022WZ20}. 
\mtrevise{
They are designed to effectively perform graph-related prediction tasks, and have achieved state-of-the-art performance in a diverse array of applications such as e-commence networks~\cite{WangHZZZL18}, anomaly detection~\cite{Wang0GYL021,DengH21,WanLWW21}, and healthcare~\cite{CaiWLL19,GopinathDL20,LiuTLW20,ManoochehriN20}. 
}
%
\mtrevise{
Because of their increasing popularity, many Machine Learning as a Service (MLaaS) platforms offered by cloud service providers have integrated GNN development tools for conveniently launching GNN services (e.g., AWS integrated DGL~\cite{adeshina_2020}, Microsoft Azure incorporated Spektral~\cite{balakreshnan_2021}, and Google Vertex AI used Neo4j~\cite{lackey_2022}). 
}
%
Despite the convenience and low cost of model deployment, graph-based MLaaS is also facing critical security concerns. 

Among others, adversarial attacks are known to be able to deduce the performance of GNNs. 
In the context of locally deployed GNNs, the limited attack surface has directed the majority of research towards the protection of these networks against \textit{data-centric attacks}~\cite{ZugnerAG19,0001DM20,sun2020non,LiuSZ0H19,Wu0TDLZ19,ZhangWY0WYP21} (e.g., evasion attacks that alter the inference graph or poisoning attacks that modify the training graph). 
Currently, \textit{model-centric attacks}~\cite{LiuWLX17,RakinHF19,YaoRF20} (e.g., bit-flipping that manipulates the GNN model parameters) are less significant and can often be mitigated through local verifying and careful monitoring of the model inference process. 

\mainmtrevise{
Nonetheless, the landscape shifts when GNNs are deployed to MLaaS platforms. 
Firstly, outsourcing GNNs to the cloud encounters practical attack surfaces that largely amplify the risk of model-centric attacks. 
Specifically, MLaaS stores GNN models in their cloud storage systems~\cite{adeshina_2020}, making them susceptible to long-standing threats inherent to cloud computing, such as misconfiguaration~\cite{MehtaB0BMAABK20,LiLLPZJW18}, insufficient access management~\cite{Kost_2023, McKeon_2023, Southwick}, and Bit-flip attacks~\cite{LiuWLX17,RakinHF19,YaoRF20}, thereby rendering model-centric attacks more applicable. 
Furthermore, GNNs located within cloud storage environments are stripped of those local protective measures that require direct access to the deployed models~\cite{abs-2006-08900,JiangZLTL19}. 
Due to security concerns, the model owner typically lacks such access within the bounds of MLaaS. 
}


A real-life illustration of this issue can be found in the collaboration between the American Heart Association and Amazon for cardiovascular disease~\cite{Heath_2016}. 
GNNs, designed to analyze proteomic data, can be deployed through Amazon's MLaaS platform, named SageMaker, and stored in the cloud within an S3 bucket as its MLaaS pipeline~\cite{adeshina_2020}. 
Unfortunately, the S3 storage system has faced both external and internal threats~\cite{PapernotMSW16,GhodsiGG17,team_2021,osborne_2021,bradford_2020,goud_2017,PriebeMKESKP14, CableGID21}. 
If an attacker were to manipulate the GNN models during the model serving process (e.g., before loading from S3), it could lead to incorrect assessments of cardiovascular disease risks or inappropriate responses to treatments and undermine the accuracy of diagnoses and treatment plans~\cite{Heath_2016}.

\begingroup\renewcommand\thefootnote{\textsuperscript{1}This work was done when Bang Wu was at Monash University}
\footnotetext{}
\endgroup

\noindent \textbf{Research Gap}. While the security risk associated with GNN models stored within MLaaS is apparent, current studies are inadequate to address them. 
\mainmtrevise{
Traditional software security solutions for integrity verification (e.g., cryptographic digests) require the verifier to access and control the data directly and validate the digests during runtime, which is not applicable within MLaaS. 
In addition, these verification methods done by the MLaaS servers assume that MLaaS is trusted, which does not align with our above considerations regarding non-trusted cloud environments.} 
Moreover, existing defenses against adversarial attacks in GNNs~\cite{abs-2006-08900,JiangZLTL19} are used to protect locally deployed GNNs against data-centric attacks, rendering them ill-equipped to counter model-centric attacks in the untrusted cloud environments.


To fill this gap, our study presents the first systematic investigation of protecting GNNs deployed in MLaaS environments against model-centric attacks through verification of model integrity during GNN serving. 
We consider GNNs in node classification tasks as the case study, i.e., a basic yet widely-adopted learning task for GNNs~\cite{GNNBook2022}. 
We employ a query-based verification mechanism that requires only query access to the targeted GNN model during the serving phase.
The core concept involves probing the deployed GNN models with a series of carefully designed node queries and analyzing their responses to assess the integrity of the GNN model. 
Responses that do not match the expected predictions indicate that the integrity of the GNN models has been compromised.

\noindent \textbf{Challenges.} Developing a comprehensive integrity verification method to protect GNNs in MLaaS environments against model-centric attacks presents a nontrivial task. 
First, outsourcing of GNNs via MLaaS introduces specific challenges. 
\textbf{C1)} \textit{Lack of Direct Access}: The interaction between the verifier and the server is limited through APIs provided by MLaaS, meaning that the verifier may not have access to the exact model parameters or internal representations during model inference. 
Thus, our verification can only rely on the basic APIs for GNN serving (i.e., prediction APIs). 
\textbf{C2)} \textit{Diversity of Query APIs}: These prediction APIs also vary across different GNN settings. 
Depending on whether GNNs operate in transductive or inductive settings, each setting dictates different inference processes, leading MLaaS to publish diverse query APIs during service (e.g., queries to transductive GNNs are typically restricted to node IDs, while those to inductive GNNs may involve new inference graphs). 
Therefore, our verification should be considerate of and adaptable to the varying settings of GNNs. 

Furthermore, the verification of GNN models also introduces unique challenges, revealing that existing studies on DNNs may not be applicable to GNNs. 
\textbf{C3)} \textit{Unique Model Architecture}: GNN models typically have fewer layers (e.g., two layers to prevent over-smoothing~\cite{KipfW17,LiHW18}) compared to other DNNs (e.g., CNNs in computer vision). 
This distinctive structure affects the activation of neurons by queries and introduces diverse focuses on verification design strategies. 
\textbf{C4)} \textit{Graph-Structured Inputs}: Unlike DNNs that operate on Euclidean data, GNNs analyze graph-structured data, where a prediction is influenced not only by its node features but also by the graph structure and its neighbors. 
Thus, our verification needs to accommodate the graph structure. 
In summary, our Research Question is: \textit{``How to design a comprehensive GNN integrity verification schema that addresses the limited and varying capabilities of MLaaS verifiers, while also accounting for the unique characteristics of GNNs?"}

\noindent \textbf{Our Contributions.} 
To address these challenges, we make the following technical contributions. 
Firstly, we direct our attention to the model-centric attacks problem for the GNNs deployed within MLaaS, and propose a comprehensive integrity verification framework against these attacks. 
Our approach adopts a query-based integrity verification workflow, which is suitable for limited model access in MLaaS (\textbf{C1}). Additionally, to adaptively deal with different query APIs (\textbf{C2}), we delineate four distinct verification types, each accommodating differing access to the deployed model under transductive/inductive settings, as well as full/limited verifier's knowledge of the GNN models. 

Secondly, recognizing a gap in existing research, we introduce new designs to generate verification queries for each of our proposed verification types. 
Within the transductive setting, where queries are limited to node IDs, we generate queries by selecting nodes that more effectively fingerprint the targeted GNN model. 
Two fingerprint score functions are defined for verifiers with either limited or full knowledge of the model.
They are designed to gauge the influence of GNN model parameters on predictions (\textbf{C3}) related to a potential verification node and its neighbors (\textbf{C4}). 
Conversely, in the inductive setting, where queries may contain node IDs and inference graphs, we introduce two fingerprint node generation methods. 
These methods construct optimized fingerprint node features and connections among neighboring nodes, grounded in the previously defined fingerprint scores. 

In addition, we enhance our design's robustness by considering an adaptive attacker seeking to identify and bypass our verification. 
We inject randomness into the selection of nodes and the construction of fingerprints complicating attackers' efforts to locate true fingerprint nodes, and thus forcing them to respond accurately to avoid detection.

In summary:
\begin{itemize}
    \item We are the first to highlight the problem of model-centric attacks that add adversarial perturbations in GNN models within MLaaS, and introduce a comprehensive framework with a query-based integrity verification workflow, suitable for the limited model access inherent in MLaaS. 
    \item We propose four unique verification methods and corresponding novel node fingerprint generation algorithms adapting to various GNN settings (transductive/inductive) and verifier's capabilities (full/limited knowledge about the GNN model). 
    \item We further consider adaptive attackers who are aware of the verification algorithms, and mitigate this potential vulnerability by implementing randomized node fingerprinting. 
    \item Through rigorous testing on four real-world datasets against five practical model-centric attacks, our proposed methods can successfully detect all these attacks while being 2-4 times more efficient than baseline designs. 
\end{itemize}

\section{Preliminary and Threat Model}
\label{sec:background}

\subsection{Graph Neural Networks within Machine Learning as a Service}
\noindent \textbf{Node Classification via GNNs.} 
Graph Neural Networks demonstrate great success in graph analysis tasks. 
In this paper, we consider a basic yet widely adopted GNN task, i.e.,  node classification. 
A node classification GNN aims to label the nodes based on both the structure and attributes of the nodes in the graph. 
%
Generally, GNN for node classification has two different learning settings: \textit{transductive} setting and \textit{inductive} setting. 
In the former setting, a training and inference graph is used to train a GNN model. 
Namely, the training graph is also used for inference. 
In practice, this setting occurs when a graph has partially labeled nodes, and GNN is used to predict the label for the rest of the nodes. 
%
For the latter setting, the training graph of the GNN model is different from the inference graph. 
This setting is similar to the traditional learning settings where GNN learns the knowledge from a training graph and is used to predict the label of other unseen graph data.


\noindent \textbf{Machine Learning as a Service for GNNs.}
Machine Learning as a Service (MLaaS) is trending as an emerging cloud service for training and serving of machine learning models. 
It enables model owners to build (\textit{Training Service}) and deploy (\textit{Serving Service}) their model and learning applications to model users conveniently and automatically. 
%
%
A notable example is the SageMaker framework provided by Amazon, which has integrated a popular graph learning tool (i.e. DGL~\cite{dgl_2020}) for the development and deployment of GNNs~\cite{simon_2019}.
%
%
In this paper, we focus on the Serving Service, and consider that model owners first upload their locally trained GNN model to the cloud storage or specify the location of an online GNN model.
Then, they create and configure the Hosting Service with an HTTPS endpoint. The Hosting Service will create a serving instance that loads the GNN model from the cloud storage bucket and launches the endpoint to host the model prediction. 
This endpoint is persistent and will remain active, so model users can make instantaneous prediction queries at any time.


\subsection{Threat Models}
\label{subsec:attackconverage} 
In this study, we specifically focus on model-centric attacks. These attacks pose a significant yet often not addressed threat in the scenario where GNNs are utilized within the framework of MLaaS.
Unlike data-centric adversarial attacks, which add perturbations to graph data, model-centric attacks aim to degrade GNN performance by compromising the integrity of GNN models, ultimately leading to inaccurate predictions. 

%
\noindent \textbf{Attack Surfaces.} As an emerging service launched by public cloud service providers, MLaaS naturally faces the same threats of cloud computing, such as misconfiguration~\cite{MehtaB0BMAABK20,LiLLPZJW18}, insufficient identity and access management~\cite{JiaXZ0Z20,PanY18}, insider threat~\cite{HomoliakTGEO19,LeKZH18}, insecure interfaces and APIs~\cite{Herardian19,HetzeltRBMS21}. 
We deem that these long-standing threats create practical attack surfaces for adversarial perturbations over GNN models. 
For example, MLaaS stores the GNN models in cloud storage, which has been demonstrated with large attack surfaces~\cite{CableGID21,GasibaALP21,Khan0SVKUZ21}. 
Specifically, an AWS S3 bucket can be accessed and overwritten by unauthorized users due to misconfigurations~\cite{team_2021,goud_2017} or exploited by an internal malicious user~\cite{osborne_2021,bradford_2020}. 
Such issues allow attackers to manipulate GNN models before being loaded by the inference instance of MLaaS. 

\noindent \textbf{Attack Coverage.} 
In this paper, we discuss two practical model-centric attacks compromising GNN model integrity. 

\noindent \textit{- Fault Injections to Model Parameters.} 
The attacker injects faults into a few bits of the GNN model parameter and forces the model to misclassify the nodes, also known as Bit-flip Attacks (BFAs)~\cite{LiuWLX17,RakinHF19,YaoRF20}. 
Specifically, attackers can identify the most vulnerable bits among the model parameters and modify only a few of them, but significantly impacting GNN predictions. 
\mainmtrevise{
BFAs are practical considerations in MLaaS systems that store the ML model in their cloud storage (e.g., Amazon SageMaker store GNNs in S3 bucket~\cite{adeshina_2020}), and could be exploited using various mature attack methods, such as RowHammer~\cite{YaoRF20,YaglikciLOOPPHK22}, VFS~\cite{BoutrosHPB20}, and clock glitching~\cite{LiuCZL20}.
}
We provide a detailed construction of BFAs on GNNs in Appendix~\ref{app:bfa}, as existing DFAs target on DNNs.

\noindent \textit{- Model Replacement by Poisoned Models.} 
Apart from only modifying a few bits, we assume another attacker who can replace the entire model with a pre-constructed poisoned one~\cite{Gupta021,SunWTHH20}. 
%
%
Specifically, we consider that the replaced models are generated through poisoning attacks, which is one of the well-studied adversarial attacks targeting GNN models and has been shown to cause severe mispredictions during GNN inference~\cite{Gupta021,SunWTHH20}.  
\mainmtrevise{
Model replacement is also a practical concern because ML models are stored in cloud storage. 
Attackers could replace models before they are loaded into the MLaaS serving instance due to well-known issues in cloud storage, such as misconfigurations~\cite{team_2021,goud_2017} and insecure APIs~\cite{Herardian19,HetzeltRBMS21}.
}
We formalize poisoning attacks in Appendix~\ref{app:poisoning_attack_definition}. 

Furthermore, we acknowledge the threat of an advanced attacker who understands the deployed defense strategies. 
Thus, we also consider an adaptive attacker who attempts to identify and bypass potential integrity verification. 
We elaborate on such adaptive attacks later in Section~\ref{sec:adaptive_design}. 

\noindent \textbf{Remark.} 
There may exist other types of threats which are out of scope in this work. 
For example, rather than targeting the serving services, the attacker attempts to compromise the MLaaS training services. 
%
Note that model-centric attacks targeting GNN models during the training phase can be trivially verified by downloading and validating the learned model on the local side.  
%
In addition, we do not focus on data-centric attacks that compromise graph data integrity in this work, as they can be mitigated by well-studied orthogonal methods. 
%
For example, poisoning/backdoor attacks can be addressed by robust training, i.e., introducing an attention mechanism to identify adversarial perturbations and decrease their weights during the aggregation process~\cite{ZhangZ20}. 
Similarly, evasion attacks can be mitigated through adversarial training~\cite{FengHTC21,JinZ20}, which makes the GNN still produce correct predictions even when the input inference graphs have been manipulated adversarially. 
\section{Queried-based Verification for GNN within MLaaS}
\label{sec:problem}
While threats of model-centric attacks targeting GNN models within MLaaS evolve, existing approaches are often either inapplicable to GNNs or unworkable within the specific constraints of MLaaS. 
Recognizing these limitations, we aim to verify the integrity of GNN models to defend against these emerging attacks. 
In the following section, we begin by defining our design objectives that aim to counter these threats, with particular attention to the MLaaS application scenario. 
We then present our four query-based verification methods, delineating their verifier's capabilities, and conclude with a detailed outline of the workflow underpinning our query-based design.

\subsection{Design Goals}
\label{subsec:verfiertaxonomy}

To defend against the aforementioned model-centric attacks that manipulate the parameters of the GNN model, our objective is to verify the integrity of the GNN model. 
This focus shapes our design goals as follows.






\vspace{-1pt}
\begin{itemize}[leftmargin=0.3cm]
    \item \textbf{Effective.} Our design should be able to detect the adversarial perturbations on the GNN model parameters with high rates. 
    \item \textbf{Robust.} Our design should be robust to adaptive attackers who know the defense mechanism in advance and attempt to bypass it. 
\end{itemize}
\vspace{-1pt}
Additionally, taking into account the specific context of GNNs deployed within MLaaS, we further consider the following two requirements: 
\vspace{-1pt}
\begin{itemize}[leftmargin=0.3cm]
    \item \textbf{Efficient.} 
    Our design should minimize the cost of the verifier and can readily be integrated into MLaaS in a minimally intrusive manner. 
    \item \textbf{Platform-independent.} 
    Our design should be independent of GNN applications and be readily adaptive to configurations of diverse MLaaS platforms. 
\end{itemize}
\vspace{-1pt}

\subsection{Design Overview}


To achieve the four design goals, we implement a query-based approach to verify the integrity of a GNN model deployed in MLaaS during its inference period. 
Specifically, we use fingerprint nodes and their expected response to uniquely represent the target models and verify their integrity. 
We expect that a modified GNN model would yield different prediction results for these nodes. 
Consequently, the verifier can generate these fingerprints based on their knowledge of the targeted GNN model and employ them by leveraging the query API to verify the model's integrity. 

\mainmtrevise{
Note that conventional software protection solutions (e.g., cryptographic digests) are insufficient to verify the GNN model's integrity in the MLaaS scenario. 
They require the verifier to have direct access to the GNN models and to validate the digests during their inference. 
However, this is not practical for third-party verifiers in the context of MLaaS, where the cloud server stores and manages the GNN model. 
Meanwhile, since we assume that the cloud server is nontrusted, these verification methods can not be done by the MLaaS server. 
}

\mtrevise{
On the contrary, a query-based mechanism can be effective, efficient, and platform-independent in the context of MLaaS.
First, the verifier is able to detect manipulations throughout the serving process, regardless of when attacks occur. 
Second, the entire process does not require additional deployments or changes on the cloud side. 
Since only prediction APIs (the essential function of MLaaS) are used, the verification is platform-independent. 
}

\mtrevise{
In addition, while existing designs only consider that the verifier has full access to the target model, we propose different verification strategies to adaptively handle various settings for GNNs in MLaaS. Specifically, we delineate four different verification methods, considering the verifier's capabilities across three key aspects: knowledge of the targeted GNN model, knowledge of the inference graph, and access to the deployed GNN model (as detailed in Table~\ref{tab:verifier_taxonomy}). In the following, we provide comprehensive explanations for these capabilities.
}


\noindent \textbf{Verifier's Knowledge of the GNN model.}
The verifier can have different knowledge of the targeted GNN model before being deployed. In practice we consider the following two cases. 

\noindent \textit{- Full Knowledge to the Targeted GNN Model.} In this case, the verifier has full knowledge of the GNN model, including the model parameter, model architecture, the gradient, and the internal representation for any local predictions before it is deployed in MLaaS. In practice, this verifier can be the model owner who wants to protect their model, or an entity fully authorized by the owner. 

\noindent \textit{- Limited Knowledge to the Targeted GNN Model.}   
In this case, the verifier does not have access to the GNN parameters and can only query the target GNN model and obtain the response. Nevertheless, the model owner is willing to support verification, and provide limited knowledge, e.g., node embeddings and prediction posteriors generated by the model inference. In practice, the verifier can be a third-party service, and the model owner does not want to share their well-trained model (IP of the owner) directly. 

\noindent \textbf{Verifier's Knowledge of the Inference Graph.}
\label{sec:problem_graphaccess}
In addition to the verifier's access to the target GNN models, we further assume the verifier's knowledge corresponding to the inference graph of the targeted GNN to facilitate verification. We also present this in the context of both transductive and inductive settings. 

\noindent \textit{- Transductive Setting.} In this setting, the verifier is provided with a small set of nodes and their neighboring subgraph within the inference (also training) graph. The intuition is that training data are often considered private data of the model owner, so they prefer to provide only a limited amount of their training graph to the verifier. 
In practice, the owner may designate a small subset of nodes specifically for security purposes, e.g., setting some nodes in a network graph as security appliances to mitigate security threats~\cite{ModiPBPPR13}. Note that if the verifier is the model owner, they will not be constrained. 

\noindent \textit{- Inductive Setting.}  In this setting, the verifier knows the prediction task of the targeted GNNs and can gather shadow graph data corresponding to this task. 
The verifier gains full access to these shadow graphs and can use them for querying. 
This is a common and practical assumption considered in previous work~\cite{ShokriSSS17,WuYPY22}. 
For example, the shadow graph could be an earlier version of the training graph provided by the model owner, or a graph that shares similar characteristics with the target GNN task~\cite{ShokriSSS17}. 


\begin{table}[t]
\centering
\footnotesize
\caption{Verification Taxonomy}
\vspace{-8pt}
\begin{tabular}{@{}cccc@{}}
\toprule
\multirow{2}{*}{Verifier}                                           & \multicolumn{2}{c}{Before Deployment} & After Deployment \\ \cmidrule(l){2-4} 
                                                                    & Targeted GNNs    & Inference Graph    & Query API        \\ \midrule
\begin{tabular}[c]{@{}c@{}}Transductive-F\end{tabular} & Full Access      & Subgraph           & Node ID          \\ \midrule
\begin{tabular}[c]{@{}c@{}}Transductive-L\end{tabular} & Limited Access     & Subgraph           & Node ID          \\ \midrule
\begin{tabular}[c]{@{}c@{}}Inductive-F\end{tabular}    & Full Access      & Shadow Graph       & Node ID, Graph   \\ \midrule
\begin{tabular}[c]{@{}c@{}}Inductive-L\end{tabular}    & Limited Access     & Shadow Graph       & Node ID, Graph   \\ \bottomrule
\end{tabular}
\label{tab:verifier_taxonomy}
\vspace{-10pt}
\end{table}

\noindent \textbf{Verifier's Access to the Deployed GNN Models.}
\label{sec:problem_queryaccess}
After the targeted GNN model is deployed, we assume that the verifier can access only the prediction APIs as the normal GNN model users. 
%
We deem that this is a practical setting due to security and cost-effective considerations.
Only leveraging prediction APIs will make the verifier hard to be detected and bypassed, and will also not introduce additional development and implementation overhead at MLaaS. 
%

\noindent \textit{- Queries to the Prediction APIs.}
The GNN prediction APIs support different formats of prediction queries depending on the GNN learning settings for node classification, i.e., transductive setting and inductive setting. 
\textbf{1)} \textit{Transductive Setting.} In this setting, the queries sent by the verifier consist only of the node ID. Since the training and inference graph is used to train the targeted GNN, the graph will not be modified during model serving. Specifically, the model owner uploads their GNN model and the training graph (aka the inference graph for predictions) to the cloud when they create the serving instance. During the serving period, users can directly issue the node ID for the serving instance to perform inference in the training graph, and respond to a corresponding label to that node. 
%
\textbf{2)} \textit{Inductive Setting.} In this setting, queries sent by the verifier consist of both the node ID and the graph used for inference. 
Differently from the transductive setting, the graphs used for inductive GNNs' training and inference are different. 
Therefore, MLaaS provides APIs for users to query their inference graph during serving. 
Specifically, the model owner uploads only their GNN model to the cloud when creating the serving instance. 
During the serving period, users upload their inference graph and its node IDs for predictions. 
The serving instance then performs inference in the inference graph, and responds to the corresponding labels of the node IDs, respectively.

\noindent \textit{- Responses from the Prediction APIs.}
%
We assume that the prediction APIs only respond to the users with the final prediction label. 
In MLaaS applications, predicting the label is the most basic function provided by the prediction APIs. 
Therefore, this is the strictest setting for the verifier, which provides the minimum information about the model inference. 
It is also consistent with recent work on machine learning security~\cite{ChangRXHZC0H20,WuYPY22}. 
%
%
We are aware that, if the response is the confidence vector, the verifier can easily verify the integrity by issuing a query and comparing it to the pre-recorded confidence vector. If the confidence scores do not match, the GNN model can be considered modified.
\label{sec:method}
\begin{figure}
    \centering
    \includegraphics[width=0.48\textwidth]{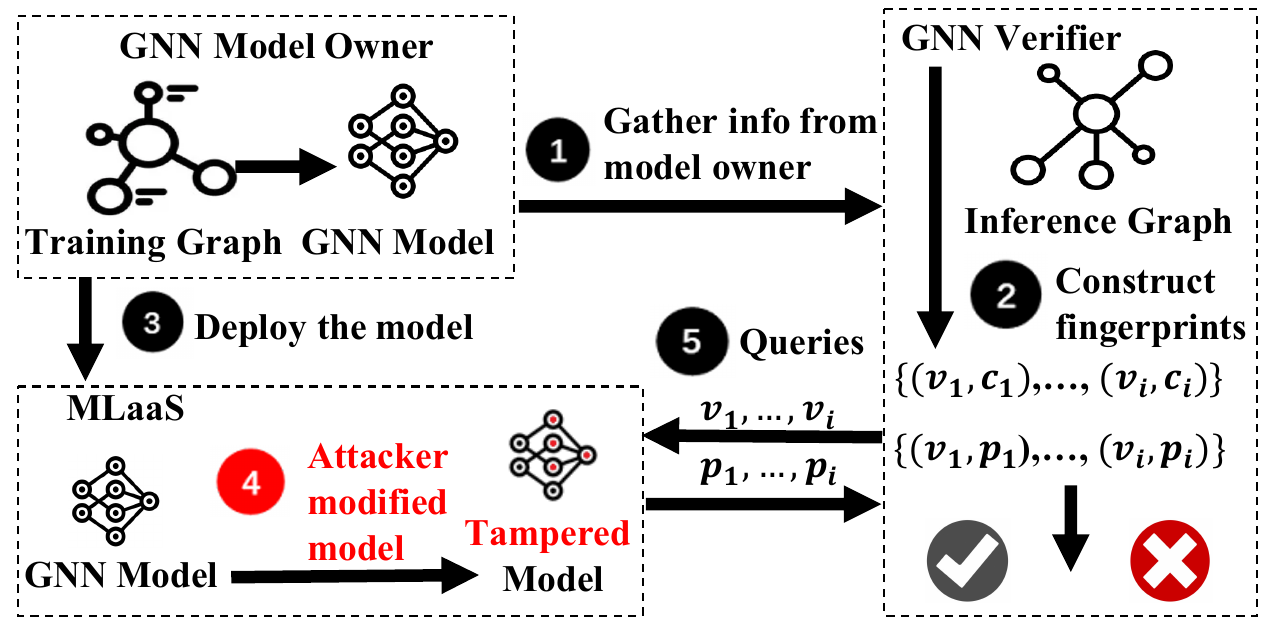}
    \caption{Overview of our verification mechanism. We target an MLaaS that provides a GNN model serving service to users through prediction APIs. A verifier is able to check the integrity of the GNN model serving by querying the APIs as a normal user. }
    \label{fig:architecture}
    \vspace{-10pt}
\end{figure}

\subsection{Design Workflow}
\label{subsec:design_overview}
For all four verification methods, we devise a universal query-based process that utilizes the aforementioned knowledge and is tailored to fit the context of MLaaS. 
\mtrevise{
In particular, our design consists of two stages: \textit{Offline Phase}, which generates node fingerprints based on the targeted GNN model locally at the verfier side (Step 1 \& 2 in Fig.~\ref{fig:architecture}) and \textit{Online Phase}, which uses node fingerprints for verification during inference (Step 3 \& 5 in Fig.~\ref{fig:architecture}).  
The attack takes place at Step 4, which occurs during the online phase of our verification process and after the GNN deployment to the MLaaS server. 
}

\textit{Offline phase} occurs before the GNN models are deployed on the MLaaS server. 
As mentioned, we consider that the model owners have developed a well-trained GNN model locally or through the MLaaS training service. 
Then, they authorize a verifier to construct the fingerprints. 
The detailed steps contained in the offline phase are as follows: 
\begin{enumerate}[leftmargin=0.5cm]
    \item The model owner provides either full or limited knowledge ($\mathcal{K}_f = \mathcal{K}_{f-F}$ or $\mathcal{K}_{f-L}$) to the targeted GNN model $f_{\theta}()$ to the verifier. 
    \item The verifier performs a fingerprint generation mechanism $\mathcal{V}(.)$ to construct a set of node fingerprints $\mathcal{F} = \{(v_1,c_1),(v_2,c_2),...,(v_i,c_i)\}$ (as well as an inference graph $G$ when in inductive settings) based on the knowledge of the inference graph $\mathcal{K}_G$, where $\mathcal{F}=\mathcal{V}(\mathcal{K}_f, \mathcal{K}_G)$.  
\end{enumerate}

\textit{Online phase} occurs after the GNN models are deployed on the MLaaS server. 
Specifically, the fingerprints constructed in the offline phase can be used during inference. 
The detailed steps contained in the online phase are as follows: 
\begin{enumerate}[leftmargin=0.5cm]
\setcounter{enumi}{2}
    \item The model owner deploys their GNN model $f_{\theta}$ via MLaaS. 
    \item The attacker could manipulate the benign GNN model $f_{\theta}$ as an adversarial model $f_{\theta'}$. 
    \item The verifier queries the prediction APIs by the fingerprint nodes $\{v_1,v_2,...,v_i\}$ (as well as an inference graph $G$ when in inductive settings) to obtain the response labels $\{p_1,p_2,...,p_i\}$, and then compares with their pregenerated labels $\{c_1,c_2,c_3\}$ to output the verification result $b = \prod_{j=1}^{i}\mathbbm{1}(c_j=p_j)$. Only if $b$ is equal to 1, the verification is carried out.
\end{enumerate}
\section{Node Fingerprint Generation Algorithms} 
In the aforementioned query-based verification mechanism, the essential part of the design process is the generation of node fingerprints tailored to the targeted model. 
As noted previously, queries issued by the verifier differ between transductive and inductive settings. 
Consequently, in this section, we will separately introduce our fingerprint generation methods, first addressing transductive scenarios, and subsequently inductive ones. 
Within each node fingerprint design, we will also consider different levels of knowledge regarding the model, i.e., full and limited access to the targeted GNNs. 
\subsection{Transductive Node Fingerprint Generation}
\label{sec:transductive_methods}
As outlined in Section~\ref{sec:background}, both the training and inference graphs are identical in the transductive settings. 
Consequently, MLaaS will not offer APIs for updating the graph during inference periods. 
As a result, queries to the targeted GNNs will exclusively contain node IDs, where fingerprints will comprise tuples consisting of node IDs and corresponding labels. 
To enable such queries to fingerprint the targeted GNN model, we propose identifying nodes that exhibit sensitivity to the GNN model. 
Specifically, alterations in model parameters will also induce changes in node predictions. 
Therefore, these pairs of nodes and labels can potentially be used as node fingerprints for GNN models.

\subsubsection{Transductive Fingerprinting}
Considering the capabilities presented in Section~\ref{sec:problem_queryaccess} (where the verifier can issue queries on a small subset of nodes, denoted as $V_c$, supplied by the model owner), a naive verification approach would involve using all these nodes as fingerprints, issuing queries for each, obtaining responses, and comparing them against the corresponding labels. 
Nevertheless, in practice, issuing queries to the target GNNs can be expensive in terms of cost (MLaaS charges based on the number of queries) and time (processing large quantities of queries and responses introduces latency). 
Therefore, we propose to minimize verification efforts by selecting the smallest possible set of nodes, denoted as $V_f$, from these candidates $V_c$ capable of detecting model modifications. 
Here, we formalize the transductive fingerprint generation problem as follows:
\begin{definition}[Transductive Node Fingerprint Generation] 
Given a set of fingerprint node candidates $V_c$ among a target graph $G=(V,E,X)$ where $V_c \subset V$, and a target GNN model $f_\theta$ with parameters $\theta$ trained on $G$, the verifier proposes to find the minimal fingerprint node set $V_f \subset V_c$, such that there exists at least a $v_j \in V_f$ which satisfies  $p_j \neq c_j$, where $p_j = \arg \max f_{\theta'}(G,v_j)$, $c_j = \arg \max f_\theta(G,v_j)$, and $\theta'$ is adversarial perturbed from $\theta$.
\end{definition}

To solve this problem, two possible strategies can be considered: (1) selecting fingerprint nodes that activate as many neurons of the model as possible, ensuring that a few nodes can cover the neurons corresponding to the modified model parameters and reflect on their predictions; and (2) selecting fingerprint nodes whose predictions are sensitive to the parameter modification, thereby increasing the sensitivity of each queried node to the model parameter modification and enhancing the likelihood of detecting the modification. 

Our proposed design resorts to the second strategy. 
In the context of GNNs, model architectures are often relatively simple, characterized by a small number of layers (e.g., only $2$ layers) and hidden features (e.g., $16$ hidden features used in our experiments). 
It is observed that only a few nodes (e.g., $2-3$ nodes) can activate most of the neurons in GNNs. 
Namely, using additional nodes beyond the initial $2$ or $3$ selected nodes does not significantly increase neuron activation. 
Thus, this activation-based selection strategy is similar to a random selection and becomes less effective. 
Consequently, our node fingerprint generation process focuses more on examining how a node's prediction changes when the GNN model is subjected to adversarial modifications. 
\subsubsection{Fingerprint Score Functions}
In this regard, we formally introduce the fingerprint score function, which quantifies the susceptibility of a node's prediction to model perturbations for generating fingerprints. 
Considering two types of verifiers (\textit{Transductive-F} and \textit{Transductive-L}) with either full or limited knowledge of the targeted model, we define fingerprint score functions for them respectively. 

\noindent \textbf{Transductive-F Verification.} 
To assess how the predictions of our fingerprinting nodes change when manipulation of the model parameter occurs, the fingerprint score function of a node can be defined as follows:
\begin{equation}
    \Phi_{i} \propto Pr(y_i \neq  y'_i),
\end{equation}
where $y_i$ is the prediction result on the clean model whose parameter is  ${\theta}$, $y'_i$ is the prediction result on the modified model whose parameter is ${\theta}'={\theta}+\Delta_{\theta}$, and $\Delta_{\theta}$ are the adversarial perturbations. 

Since the prediction result $y_i$ is non-linear to the model output $f_{{\theta}'}(G, v_i)$, we approximate the difference between the final prediction labels by the L2 distance between their output confidence vectors, as follows (detailed proof can be found in Appendix~\ref{app:fingerprinting_proof}):
\begin{equation}
\begin{aligned}
& \Phi_{i} \propto 
\left \|{\left(\frac{\partial f_{{\theta}}(G, v_i)}{\partial {\theta}}\right)}^T \Delta_{\theta}\right \|_2, \\
\end{aligned}
\end{equation}
In general, $\Delta_{\theta}$ is a small perturbation as attackers often tend to be less noticeable. 
Therefore, we disregard this higher-order term and define our fingerprint score:
\begin{definition}[Fingerprint Score for Transductive-F] Given a GNN model $f_{{\theta}}(G,v_i)$ with $N$ model parameters $\theta=\{\theta_j, j \in [1,N]\}$ trained on the graph $G=(V,E,X)$, the fingerprint score for \textit{Transductive-F} of a node $v_i \in V_c$ corresponding to the perturbation at ${\theta}$ is defined as follows:
%
\begin{equation}
    \Phi_{i} = \sum_{\theta_j \in {\theta}}\left \|\frac{\partial \mathcal{L}(f_{{\theta}}(G,v_i),y_i)}{\partial \theta_j}\right \|_2,
    \label{eqt:grad-sensitive_values}
\end{equation}
\end{definition}
where $\mathcal{L}(.)$ represents the distance between the node prediction and its prediction label $y_i$. 
Since the verifier is assumed to have full knowledge of the target GNN model, by performing backward propagation as model training, the verifier can easily calculate the gradient in the above equation and obtain the final fingerprint scores. 

\mtrevise{
Note that in calculating the gradient mentioned above, we do not compute scores in Eq.~\ref{eqt:grad-sensitive_values} for every node. 
Instead, we calculate scores for a selected set of candidate nodes $V_c$ and choose fingerprints from among them. 
As a result, the computational cost of our design is confined within a range determined by the size of $V_c$. 
In practice, the pool of candidates can be adaptively expanded to include all nodes or narrowed to include any subset of nodes in the graph. 
}

\noindent \textbf{Transductive-L Verification.} 
In this case, the verifier cannot have full knowledge of the GNN models, which prevents them from calculating the gradients. 
To address the problem, we propose another node generation method that does not rely on the gradient. 
Because our verification is query-based and correlated with the final prediction, we utilize the prediction scores to evaluate how parameter perturbations may affect the final inference results. 
In particular, fingerprinting nodes with a lower level of confidence in their final prediction will be more likely to be affected by the perturbation. 
Therefore, the fingerprint score of a fingerprinting node for \textit{Transductive-L} can follow: 
\begin{equation}
    \Phi_{i} \propto \mathcal{D}(f_{{\theta}}(G, v_i),\hat{y}_i),
\end{equation}
where $f_{{\theta}}(G, v_i)$ represents the output of the GNN classifier for the node $v_i$, $\hat{y}_i$ represents its one-hot prediction label vector, and $\mathcal{D}(.)$ represents the distance between the current confidence vector and the one-hot prediction label vector of the node $v_i \in V$. 
%
Consequently, we define our fingerprint score of a node for \textit{Transductive-L} as follows:
\begin{definition}[Fingerprint Score for Transductive-L] Given a GNN model $f_{{\theta}}(G,v_i)$ with $N$ model parameters ${\theta}=\{\theta_i, i \in \{1,2,...,N\}\}$ trained on the graph $G=(V,E,X)$, the fingerprint score for Transductive-L of the node $v_i \in V_c$ corresponding to the perturbations is defined as follows:
    \begin{equation}
    \label{eqt:adv-sensitive_values}
        \Phi_{i} = \mathcal{D}(f_{{\theta}}(G, v_i),\hat{y}_i)
    \end{equation}
\end{definition}

It should be noted that the above fingerprint scores are calculated only based on the prediction result of a node, whose calculation cost is $O(1)$, making it efficient for our verification. 
In practice, to further speed up the calculation of the above fingerprint scores, we replace Eq.~\ref{eqt:adv-sensitive_values} with: 
\begin{equation}
    \Phi_{i} = 1-\mathrm{Softmax}(f_{{\theta}}(G, v_i))[y_i],
\end{equation}
where $\mathrm{Softmax}(\cdot)$ outputs a normalized probability vector of a specific node prediction with $c$ dimensions, and $c$ represents the number of classes for the node classification tasks. 

Based on the above definitions, the fingerprint scores represent how the model weights can affect the final prediction loss. 
With higher fingerprint scores, the prediction results of such nodes can be affected more easily, and they can be considered as fingerprinting nodes for verification during model serving. 
%
As a result, the verifier can calculate the fingerprint scores for a set of candidate nodes (e.g., nodes owned by the model owner that can be specifically used for verification), choose the nodes that will be easily impacted by the perturbation applied to the GNN model parameters, and construct a node fingerprint as a tuple consisting of them and their corresponding label. 

\noindent \textbf{Remark.} 
Note that our transductive fingerprinting approach differs from the sensitive sample fingerprints presented in prior work by He et al.~\cite{HeZL19}. 
In their study, the assumption is that the verifier has the ability to issue queries containing adaptive input images. 
On the contrary, our transductive fingerprinting is limited to issuing queries with node IDs only, resulting in a significantly more limited capability. 
Additionally, their work only considers cases where the verifier has full knowledge of the targeted model. 
However, in our approach, we propose two fingerprint generation methods, taking into account that the verifier may also have only limited knowledge of the target model, which cannot be solved by their method.
Furthermore, they used a Maximum Active Neuron Cover (MANC) Sample Selection technique upon obtaining a set of fingerprint candidates, which selects fingerprints based on neuron activation rather than their sensitivity to model parameters. 
As discussed previously, relying solely on neuron activation will be less effective in the context of GNNs. 
In Section~\ref{sec:exp_trans}, we compare our design with their approach in our evaluation, further highlighting the distinctions between the two methods.

\subsection{Inductive Node Fingerprint Generation}
\label{sec:inductive_methods}
Contrary to the transductive setting, the training and inference graphs for inductive GNNs are distinct. 
In this scenario, MLaaS provides APIs that allow GNN users to update the inference graph during the serving periods. 
Consequently, the query sent by the verifier comprises both the node IDs and the inference graph, indicating that our proposed fingerprints should also include a graph, the fingerprint node IDs, and their corresponding labels. 
Thus, our design should construct an inference graph with several fingerprint nodes, anticipating that any modifications to the model parameters will alter the predictions on the fingerprint nodes within our constructed graph. 

\subsubsection{Inductive Fingerprinting}
Our above inductive fingerprint generation can be divided into two steps: 1) constructing an inference graph and 2) selecting the fingerprint node indices. 
When constructing the graph in Step 1, it is essential to ensure that the constructed inference graph resembles the ordinary inference graph of the target GNN tasks to prevent adversaries from being aware of and bypassing our verification. 
Therefore, the verifier will gather or construct a shadow graph as the ordinary inference graph (refer to the capabilities discussed in Section~\ref{sec:problem_graphaccess}). 
After that, we can follow a similar strategy as in the transductive setting to select the node indices for Step 2. 
The problem can be formalized as follows.
\begin{definition}[Inductive Fingerprinting Generation] 
Given a shadow graph $\hat{G}=(\hat{V},\hat{E},\hat{X})$, and a target GNN model $f_\theta$ with parameters $\theta$ trained on $G$, the verifier proposes to find a minimal fingerprint node set $V_f \subset \hat{V}$, such that there exists at least a $v_j \in V_f$ which satisfies  $p_j \neq c_j$, where $p_j = \arg \max f_{\theta'}(\hat{G},v_j)$, $c_j = \arg \max f_\theta(\hat{G},v_j)$, and $\theta'$ is adversarial perturbed from $\theta$.
\end{definition}

Additionally, rather than directly using the gathered shadow graph, we propose further perturbing the shadow graph to increase the effectiveness of fingerprinting. 
This can enhance the sensitivity scores of the fingerprint nodes without significantly altering the graph, preventing adversaries from bypassing it. 
In our design, we propose to perturb the graph structure and node features corresponding to these fingerprint nodes, as their sensitivity scores are primarily impacted by these components. 
Note that, we do not consider injecting or deleting nodes from the graph, as adding or removing nodes can be regarded as using a different gathered shadow graph.

\subsubsection{Fingerprint Nodes Construction}
We now discuss how to apply perturbations to the fingerprint nodes to increase their sensitivity scores. 
Based on the verifier's knowledge of the targeted GNN models, we design different strategies for both \textit{Inductive-F} and  \textit{Inductive-L} to construct node fingerprints.



\noindent \textbf{Inductive-F Verification.} 
To increase the sensitivities of the selected fingerprinting nodes, we propose to further manipulate their attributes and connections.
%
%
Meanwhile, model performance must be preserved for quality model serving. 
Therefore, we aim to maximize the fingerprint score of the fingerprint nodes by manipulating their attributes and connections without affecting the original prediction results. 
Specifically, we formalize node construction to an optimization problem as follows:
\begin{equation}
\label{eqt:case1}
\begin{aligned}
\max_{\hat{G}'=(\hat{V},\hat{X}',\hat{E}')} \quad & \sum_{v_i \in V_f} \frac{\partial \mathcal{L}(f_{\theta}(\hat{G}',v_i),y_i)}{\partial (\hat{x}_{i}^{'},\hat{e}_{i,j}^{'})}\\
\textrm{s.t.} \quad & \forall v_k \in \hat{V}, f_{\theta}(\hat{G}',v_k)=f_{\theta}(\hat{G},v_k)\\
\end{aligned}
\end{equation}
where $\hat{x}_{i}^{'} \in \hat{X}'$ is the perturbed node attribute for node $v_i$ and $\hat{e}_{i,j}^{'} \in \hat{E}'$ is the edge connected to node $v_i$ in the perturbed graph structure. 
The right-hand side of the objective function is in a similar form to Eq.~\ref{eqt:grad-sensitive_values} but represents how the prediction of the selected nodes can be affected by the current perturbed graph. 
Furthermore, to ensure that our modification does not affect the original performance of the target model, we set a constraint such that all predictions of the nodes in the graph are not affected by the perturbation. 

Due to the discretion of the graph and even the attributes, it is nontrivial to solve the above optimization problem. 
Here, we design a greedy algorithm to increase the total fingerprint scores iteratively. 
Specifically, we first calculate the gradient of the structure and the attributes corresponding to the fingerprint scores of the nodes. 
Then, we sort all these values and add the perturbations greedily. 
After adding each perturbation, we check whether it affects the second-order neighbor of both the fingerprinting nodes and the connected nodes if perturbing the edges. 
If the perturbation leads to the misclassification of any nodes, we skip this perturbation. 
The above perturbation will not stop, until the perturbation budget is reached or adding the perturbation leads to a decrease in the fingerprint scores. 

\noindent \textbf{Inductive-L Verification.} 
We then discuss the verification strategies with limited knowledge of the targeted model. 
As mentioned above, the node located at the decision boundary of the target model is less robust, and may easily cross the boundary due to the perturbations. 
Based on this intuition, we utilize the methodology for general adversarial attacks (e.g., Nettack~\cite{ZugnerAG19}) to design the fingerprinting nodes. 
Specifically, we perturb the previously selected fingerprinting nodes by moderately increasing the prediction loss of these nodes. 
With such manipulation, the generated nodes can be closer to the decision boundary and become more sensitive to the perturbations. 
Unlike ordinary adversarial attacks, here we need to constrain the final prediction of all the nodes to be correct and make sure that our constructed nodes will not affect the GNN performance.  
Formally, our method can be represented as the optimization function:
\begin{equation}
\label{eqt:induct_blackbox}
\begin{aligned}
\max_{\hat{G}'=(\hat{V},\hat{X}',\hat{E}')} \quad & \sum_{v_i \in V_f} \mathcal{L}(f_{\theta}(\hat{G}',v_i),y_i)\\
\textrm{s.t.} \quad & \forall v_k \in \hat{V}, f_{\theta}(\hat{G}',v_k)=f_{\theta}(\hat{G},v_k)\\
\end{aligned}
\end{equation}
Compared to the optimization functions for \textit{Inductive-F} (as Eq.~\ref{eqt:case1}), the gradient generated from the target model is no longer an essential component. 
For the rest of the processes, we follow a similar approach as \textit{Inductive-F}. 
We first select a set of fingerprinting nodes and then perform our fingerprint construction method. 

\noindent \textbf{Remark.}
It is important to note that our inductive fingerprinting approach also diverges from the one presented in~\cite{HeZL19}. 
Similarly to our transductive design, our inductive approach considers both full and limited knowledge of the target GNN model, while the above work only accounts for full access and cannot deal with the verifier with limited access.
Furthermore, their sensitive samples are generated individually, with predictions based solely on the generated samples, making them more suitable for image classification. 
In contrast, within the context of GNNs, node predictions can be influenced not only by the fingerprint node features, but also by the neighboring node features and their connections. 
Aside from generating node fingerprints with their node features, our approach also considers perturbing the discrete graph structure. 
Given that GNN predictions can be significantly impacted by the graph structure, incorporating this factor into our design enhances its effectiveness. 
In Section~\ref{sec:exp_trans}, we compare the above work with ours in our evaluation, further highlighting the differences between the two methods.

\section{Mitigating Adaptive Attackers}
\label{sec:adaptive_design}
\subsection{Adaptive Attackers Knowing Node Fingerprinting Methods}
%
Our current design can effectively deal with non-adaptive adversarial manipulations of the GNN models. For practical considerations, we also assume a powerful adaptive attacker who has knowledge of the methods used to select and generate verification nodes and the size of verification node sets. 
Among others, this attacker attempts to identify our verification queries. 
 We consider that such a strong attacker can always bypass our verification process once he determines the verification queries correctly. 
They may, for example, force the MLaaS server to respond with the correct labels to those identified verification queries. 

In the context of such advanced attacks that understand our design strategies, our above design may be bypassed and become less effective. 
In the case of transductive GNNs, the attacker is able to precalculate fingerprint scores for nodes and then mark those nodes with high fingerprinting scores as potential verification nodes. 
As long as the attacker detects a query containing the IDs of these potential verification nodes, he can predict the labels of those nodes on the clean GNN model, thus circumventing our detection mechanism. 
In the case of inductive GNNs, once the attacker detects a query with an updated graph, he can also apply the calculation outlined above to the nodes in the previous version of the graph. 
Similarly, whenever the graph update is the same as the fingerprint update, these queried nodes are considered potential verification nodes and will be answered honestly.

\subsection{Randomized Node Fingerprinting}
To address these advanced attackers, we propose an enhanced design for our framework. 
Intuitively, the verification methods that we use are bypassed because they are static, whereas the fingerprint nodes derived from the specific inference graph are fixed. 
Thus, an advanced attacker can always follow the exact approach as our design and identify the fingerprint nodes that the verifier has constructed. 
Therefore, to prevent being identified, we have to introduce randomness into our design to ensure that the fingerprint nodes are not fixed and can hardly be identified. 
%
We propose strategies for introducing randomness into our fingerprint verification in inductive and transductive settings, respectively. 

\noindent \textbf{Transductive Setting.}
In this setting, rather than greedily selecting fingerprint nodes among all possible verification nodes, we first randomly select several nodes as the candidates. 
From these candidates, we select the final fingerprint nodes according to the methods we introduced above. 
Note that we conduct our fingerprint node search after random sampling. 
This not only reduces the computation cost of our search algorithm, but also makes it more difficult for attackers to identify the fingerprint nodes. 
Generally, if the fingerprint nodes are first searched and then sampled, the attackers can always bypass verification by increasing the size of their potential fingerprint node set up to the total number of the sampled candidates. 
Nevertheless, in our proposed method, they need a much larger set of candidate nodes to ensure that the true fingerprint nodes are included. 
Formally, we present Theorem~\ref{theorem:tranduct_adaptive} and perform a theoretical security analysis to prove it in Appendix~\ref{app:transduct_adaptive}.
\begin{theorem}
\label{theorem:tranduct_adaptive}
    Consider an adaptive attacker $\mathcal{A}$ following the same strategy as our verifier $\mathcal{V}_{trans}$ to generate $m_A$ fingerprint nodes in the transductive setting, the probability for $\mathcal{A}$  to include all $m_V$ fingerprint nodes generated by $\mathcal{V}_{trans}$ and bypass verification (defined as \textit{Event} $\mathscr{A}$) is
    \begin{equation}
        Pr(\mathscr{A}) \approx {(\frac{m_A}{N})}^{m_V}.
    \end{equation}
\end{theorem}
This probability becomes negligible when $m_V$ increases unless $m_A \approx N$, where $N$ is the total number of nodes on the graph.
However, if $m_A$ approaches $N$, the attacker will honestly respond to almost all queries from nodes, which makes their attack meaningless. 

%
\noindent \textbf{Inductive Setting.}
For this setting, we propose to replace the fixed node label in Eq.~\ref{eqt:induct_blackbox} with a randomly selected label and propose to minimize the prediction loss as:
\begin{equation}
    \label{eqt:inductive_adaptive}
    \begin{aligned}
        \min_{(X^{'},A^{'})} \quad & \mathcal{L}(f(X^{'},A^{'})_{i},\mathcal{R}(Y_i))\\
        \textrm{s.t.} \quad & \forall v_i \in V, f(X^{'},A^{'})_i=f(X,A)_i\\
    \end{aligned}
\end{equation}
where $R(Y_i)$ is a function that generates a random index of ${Y_i}$, where $Y_i$ is the label sets of the target graph. 
In Eq.~\ref{eqt:inductive_adaptive}, rather than misclassifying the fingerprint node to a fixed label, our optimization problem will misclassify it to a randomly selected label. 
Thus, randomness is introduced to our optimization process by forcing convergence via a randomly selected trajectory. 
Consequently, the verifier will produce different fingerprint nodes and increase the difficulty of the attacker bypassing our verification. 
Similarly, we present Theorem~\ref{theorem:induct_adaptive} for the security guarantee in an inductive setting.
\begin{theorem}
\label{theorem:induct_adaptive}
    Consider an adaptive attacker $\mathcal{A}$ following the same strategy as our inductive verifier $\mathcal{V}_{ind}$ to generate $m_A$ fingerprint nodes, the probability for $\mathcal{A}$ to include all $m_V$ fingerprint nodes generated by $\mathcal{V}_{ind}$ and bypass verification (defined as \textit{Event} $\mathscr{B}$) is
    \begin{equation}
        Pr(\mathscr{B}) \approx {(\frac{m_A}{cN})}^{m_V},
    \end{equation}
    where $c$ is the total number of classes for a node.
\end{theorem}
Note that $Pr(\mathscr{B}) \leq Pr(\mathscr{A})$ proves that our randomized node fingerprint generation for inductive settings can also effectively deal with the adaptive attacker. 

\noindent \textbf{Other Benefits.}
While being robust in dealing with an advanced attacker, our enhanced designs also satisfy the other two design goals: efficient and platform-independent. 
Both of them introduce randomness before applying our ordinary algorithm, thus, they will introduce marginal effort. 
Meanwhile, all these processes can be done on the verifier side only. 
Because they are platform-independent, it is possible for verifiers to readily update their previous verification mechanisms to the enhanced ones even after the GNN models are deployed. 
Namely, the GNN developer can directly ask the verifier to regenerate the randomized fingerprints without any upgradation on the server side. 

\mtrevise{
Note that, although introducing randomization may slightly reduce the effectiveness of the generated node fingerprinting, our proposed enhanced design can still be sufficient for verification. 
In practice, it is not essential for all fingerprints to be sensitive to perturbations.
A mismatch in one fingerprint indicates that the model has been perturbed. We observe that even after randomization, it is highly probable that at least one significant node for verification is included. 
The verifier can also expand the set of sampled nodes to ensure robust detection performance.
}

\section{Experiments and Evaluations}
\label{sec:exp}
\subsection{Experiment Setup}

        

        

        


\noindent \textbf{Datasets and Models.}
Four public datasets are utilized to evaluate the proposed attack and verification methods, namely Cora, Citeseer, Pubmed~\cite{KipfW17}, and Polblog~\cite{AdamicG05}. 
These datasets serve as benchmark standards and are extensively used in the assessment of node classification GNNs.
Cora, Citeseer, and Pubmed are citation networks in which nodes represent publications and the edges correspond to their citations. 
Each node in these networks includes attributes (i.e., keywords from the associated publication).
In contrast, Polblog is a blog dataset, where each node represents a political blog, and the edges denote links between the blogs. 
Unlike the other datasets, Polblog does not include any attributes associated with its nodes. 

In our experiments, we consider the target model to be a 2-layer graph convolution neural network model.
The number of features in the hidden layer for both GNN models is 16.
The activation function for the hidden layer is ReLU and for the output layer is softmax.
In addition, we apply a dropout layer with a dropout rate of 0.5 after the hidden layer.
We use the Adam optimizer with a learning rate of 0.02 and training epochs of 200.
The loss function of our model is negative log-likelihood loss. 
Table~\ref{tab:data_statistic} lists the statistics of the datasets. 

\begin{table}[t]
\footnotesize
    \label{tab:data_statistic}
    \centering
    \caption{GNN Training Graph Statistics}
    \vspace{-8pt}
    \begin{tabular}{c|ccc}
    \toprule
        Datasets &  Node \#  & Edge \# &  Attribute \#  \\
    \toprule
      Cora & 2708 & 5429 & 1433 \\
    \midrule
      Citeseer & 3312 & 4732 & 3703 \\
    \midrule
      Pubmed & 19717 & 44338 & 500 \\
    \midrule
      Polblogs & 1490 & 33430 & - \\
    \bottomrule
    \end{tabular}
  \vspace{-10pt}
\end{table}

\noindent \textbf{Attack Coverage.}
Two types of attacks, BFAs and Poisoning attacks (as introduced in Section~\ref{subsec:attackconverage}) are used to evaluate our design. 
%
%
%

\noindent \textit{- BFAs.} 
To evaluate how our design can detect the adversarial perturbation caused by the fault injections to the GNN model parameters, we implement Bit-flip Attacks to the GNN models by following existing attacks in DNNs~\cite{LiuWLX17,BreierHJMB018}. 
Specifically, we consider three types of bit flipping: 
\textit{BFA}, where an attacker will manipulate one randomly selected parameter among the models and flip the most significant bit of the exponent part;
\textit{BFA-L}, where the attacker tampers with the bias of the last layer of GNNs; 
\textit{BFA-F}, where the attacker tampers with the bias of the first layer of GNNs. 
Detailed settings and attack results can be found in the Appendix~\ref{app:bfa}. 
\noindent \textit{- Poisoning Attacks.} 
We also apply poisoning attacks on GNNs to evaluate our design when verifying the GNN model replaced by a poisoned model generated via adversarial attacks. 
%
Our evaluation will consider two representative attacks: Mettack~\cite{ZugnerG19} and a Random attack as a baseline attack that randomly adds edges to perturb the training graph to produce an adversarial GNN model. 
We include a detailed introduction of the poisoning attack in Appendix~\ref{app:poisoning_attack_definition}. 
%
%
%


\noindent \textbf{Baselines.}
In our study, we evaluate both transductive and inductive settings of GNNs and consider two baseline designs for each setting.

\noindent \textit{- Transductive Baselines.} 
In the transductive setting, the verification queries consist only of node IDs. We adopt the sensitive sample selection technique \textit{MANC} from~\cite{HeZL19} as the first baseline. 
MANC aims to identify samples that activate the most neurons in the target machine learning model and utilize them for verification. 
Second, we introduce the \textit{Random Node Selection} method as a baseline, where the verifier randomly selects nodes within the target graph and issues queries for verification. 
This approach represents the most fundamental baseline design for comparison.


\noindent \textit{- Inductive Baselines.}
In the case of inductive GNNs, the verification queries include an inference graph. 
We first adapt the fingerprinting generation method, sensitive-sample (denoted as \textit{SS}), for DNNs from~\cite{HeZL19} to GNNs. 
This method commences with the generation of a single sensitive sample using a gradient-based algorithm on specific nodes' attributes, followed by the application of the MANC sample selection method to choose multiple samples for verification. 
In addition, we also introduce the \textit{Ordinary Inference Graph} method as a baseline, where the verifier randomly queries nodes from an inference graph in the same domain as the target GNNs' prediction task for verification purposes. 

\noindent \textbf{Evaluation Metrics.} 
To evaluate our methods, we focus on the performance corresponding to our design goals. 

\noindent \textit{- Effectiveness.}
To demonstrate how our design can successfully detect different types of adversaries, we first define a successful detection as, when sending queries from $n$ fingerprint nodes, there exists at least one node whose label prediction from the response is different from its recorded label generated from the intact GNN. 
Then, we define the \textit{Detection Rate} (\textit{DR}) as the percentage of launched attacks successfully detected by our verification methods. 
For example, for BFAs, we first generate node fingerprinting pairs, repeat the BFAs $400$ times, and gather the number of attacks that successfully deduct the model accuracy of more than $5\%$. 
Then, we use our verification methods to detect these attacks. 
\textit{DR} is calculated by:
\begin{align*}
    \textit{DR} = \frac{\textit{\# successful detection}}{\textit{\# valid attacks}}
\end{align*}

%
\noindent \textit{- Efficiency.}
To demonstrate the efficiency of our design, we will show that our method only requires a small set of node fingerprints.
To further substantiate the efficiency of our design, we compare our approach with the baselines and introduce a metric called \textit{Query Number Improvement} (\textit{QI}) to evaluate the improvement in our design over the baseline. 
Specifically, we use $\textit{QI}_{(A,B)}$ to represent the improvement in the number of queries from verification method A to method B when both methods are capable of achieving a detection rate $100\%$. 
\textit{QI} is calculated as follows: 
\begin{align*}
    \textit{QI}_{(A,B)} = \frac{\textit{query \# for A }}{\textit{query \# for B}}
\end{align*}
We also assess the cost of deploying our design by evaluating how seamlessly it can be integrated into MLaaS. 
This assessment includes considering the number of lines of code required for implementation across different entities, as well as the time cost associated with deploying our design. 

\noindent \textit{- Robustness.} To demonstrate the effectiveness of our enhanced designs against adaptive attacks, we define the \textit{Bypass Rate} (\textit{BR}) to assess how adaptive attackers can identify our verification to bypass it. 
\textit{BR} is calculated by: 
\begin{align*}
    \textit{BR} = \frac{\textit{correctly identified verification \#}}{\textit{true verification \#}}
\end{align*}




\noindent \textit{- Platform-independence.}
We evaluate this goal of our design by reporting both the deployment cost of our methods on the verifier side and the MLaaS side. 
We show that, since there is no additional cost on the MLaaS side, our design is applicable to arbitrary types of MLaaS. 
Namely, our design can be readily applied to different MLaaS platforms. 
\subsection{Effectiveness}
\label{sec:exp_trans}

\begin{table*}[t!]
\centering
\footnotesize
\tabulinesep=1.1mm
\caption{Comparison of Single Verification Detection Rates between Our Method and Baselines for Transductive-F.}
\vspace{-8pt}
\begin{threeparttable}
\begin{tabularx}{\textwidth}{c|*{12}{>{\centering\arraybackslash}X}}
\hline
\multirow{2}{*}{} & \multicolumn{3}{c}{Cora} & \multicolumn{3}{c}{Citeseer} & \multicolumn{3}{c}{Polblog} & \multicolumn{3}{c}{Pubmed} \\ \cline{2-13}
 & Random & MANC & Ours & Random & MANC & Ours & Random & MANC & Ours & Random & MANC & Ours \\ \hline
BFA & 0.031 & 0.049 & \textbf{0.711} & 0.047 & 0.290 & \textbf{0.586} & 0.460 & 0.125 & \textbf{0.579} & 0.556 & 0.179 & \textbf{0.813} \\

BFA-F & 0.950 & \textbf{0.999} & 0.960 & 0.730 & \textbf{0.840} & 0.430 & 0.366 & \textbf{0.590} & 0.430 & 0.569 & 0.310 & \textbf{0.793} \\

BFA-L & 0.375 & 0.167 & \textbf{0.500} & 0.488 & 0.135 & \textbf{0.529} & 0.984 & 1.00 & \textbf{1.00} & 0.765 & 0.706 & \textbf{1.00} \\

Random & 0.342 & 0.002 & \textbf{0.647} & 0.410 & \textbf{0.741} & 0.412 & 0.582 & 0.701 & \textbf{0.765} & 0.998 & 0.953 & \textbf{1.00} \\

Mettack & 0.352 & 0.001 & \textbf{0.588} & 0.174 & 0.000 & \textbf{0.353} & 0.035 & 0.000 & \textbf{0.176} & 0.998 & 1.0 & \textbf{1.00} \\
\hline
\end{tabularx}
\end{threeparttable}
\label{tab:effective_trans_f}
\vspace{-10pt}
\end{table*}

\begin{table*}[t!]
\centering
\footnotesize
\tabulinesep=1.1mm
\caption{Comparison of Single Verification Detection Rates between Our Method and Baselines for Transductive-L.}
\vspace{-8pt}
\begin{threeparttable}
\begin{tabularx}{\textwidth}{c|*{12}{>{\centering\arraybackslash}X}}
\hline
\multirow{2}{*}{} & \multicolumn{3}{c}{Cora} & \multicolumn{3}{c}{Citeseer} & \multicolumn{3}{c}{Polblog} & \multicolumn{3}{c}{Pubmed} \\ \cline{2-13}
 & Random & MANC & Ours & Random & MANC & Ours & Random & MANC & Ours & Random & MANC & Ours \\ \hline
BFA & 0.031 & - & \textbf{0.982} & 0.047 & - & \textbf{0.289} & 0.460 & - & \textbf{0.763} & 0.556 & - & 0.555 \\

BFA-F & 0.952 & - & \textbf{0.810} & 0.730 & - & 0.430 & 0.600 & - & \textbf{0.937} & 0.569 & - & \textbf{0.920} \\

BFA-L & 0.375 & - & \textbf{1.00} & 0.488 & - & 0.133 & 1.00 & - & \textbf{1.00} & 0.765 & - & \textbf{0.952} \\

Random & 0.324 & - & \textbf{0.353} & 0.412 & - & \textbf{0.824} & 0.588 & - & \textbf{0.765} & 0.998 & - & \textbf{1.00} \\

Mettack & 0.353 & - & 0.598 & 0.176 & - & \textbf{0.235} & 0.023 & - & \textbf{0.118} & 0.998 & - & \textbf{1.00} \\
\hline
\end{tabularx}
\end{threeparttable}
\label{tab:effective_trans_l}
\vspace{-10pt}
\end{table*}

\begin{table*}[t!]
\centering
\footnotesize
\tabulinesep=1.1mm
\caption{Comparison of Single Verification Detection Rates between Our Method and Baselines: Ordinary Inference Graph (denoted as Ordinary), and Sensitive sample fingerprinting (denoted as SS), for Inductive-F.}
\vspace{-8pt}
\begin{threeparttable}
\begin{tabularx}{\textwidth}{c|*{12}{>{\centering\arraybackslash}X}}
\hline
\multirow{2}{*}{} & \multicolumn{3}{c}{Cora} & \multicolumn{3}{c}{Citeseer} & \multicolumn{3}{c}{Polblog} & \multicolumn{3}{c}{Pubmed} \\ \cline{2-13}
 & Ordinary & SS & Ours & Ordinary & SS & Ours & Ordinary & SS & Ours & Ordinary & SS & Ours \\ \hline
BFA & 0.031 & 0.118 & \textbf{0.667} & 0.047 & 0.003 & \textbf{0.941} & 0.460 & - & \textbf{0.800} & 0.536 & 0.199 & \textbf{0.885} \\

BFA-F & 0.901 & 0.882 & \textbf{1.00} & 0.530 & 0.529 & \textbf{0.882} & 0.366 & - & \textbf{0.395} & 0.549 & 0.313 & \textbf{0.804} \\

BFA-L & 0.327 & \textbf{0.567} & 0.382 & 0.488 & 0.513 & \textbf{0.529} & 0.984 & - & \textbf{1.00} & 0.767 & 0.716 & \textbf{1.00} \\

Random & 0.323 & 0.792 & \textbf{1.0} & 0.994 & 1.0 & \textbf{1.0} & 0.588 & - & \textbf{0.706} & 0.998 & 0.995 & \textbf{1.00} \\

Mettack & 0.561 & 1.0 & \textbf{1.0} & 0.994 & 1.0 & \textbf{1.0} & 0.035 & - & \textbf{0.176} & 0.998 & 1.0 & \textbf{1.00} \\
\hline
\end{tabularx}
\end{threeparttable}
\label{tab:effective_in_f}
\vspace{-10pt}
\end{table*}

\begin{table*}[t!]
\centering
\footnotesize
\tabulinesep=1.1mm
\caption{Comparison of Single Verification Detection Rates between Our Method and Baselines: Ordinary Inference Graph (denoted as Ordinary), and Sensitive sample fingerprinting (denoted as SS), for Inductive-F.}
\vspace{-8pt}
\begin{threeparttable}
\begin{tabularx}{\textwidth}{c|*{12}{>{\centering\arraybackslash}X}}
\hline
\multirow{2}{*}{} & \multicolumn{3}{c}{Cora} & \multicolumn{3}{c}{Citeseer} & \multicolumn{3}{c}{Polblog} & \multicolumn{3}{c}{Pubmed} \\ \cline{2-13}
 & Ordinary & SS & Ours & Ordinary & SS & Ours & Ordinary & SS & Ours & Ordinary & SS & Ours \\ \hline
BFA & 0.028 & - & \textbf{0.688} & 0.047 & - & \textbf{0.901} & 0.461 & - & \textbf{0.752} & 0.532 & - & \textbf{0.855} \\

BFA-F & 0.871 & - & \textbf{0.989} & 0.520 & - & \textbf{0.852} & 0.346 & - & \textbf{0.402} & 0.553 & - & \textbf{0.844} \\

BFA-L & 0.307 & - & \textbf{0.348} & 0.488 & - & \textbf{0.569} & 0.984 & - & \textbf{1.00} & 0.767 & - & \textbf{0.967} \\

Random & 0.313 & - & \textbf{1.00} & 0.991 & - & \textbf{1.00} & 0.587 & - & \textbf{0.746} & 0.997 & - & \textbf{1.00} \\

Mettack & 0.562 & - & \textbf{1.00} & 0.903 & - & \textbf{1.00} & 0.035 & - & \textbf{0.216} & 0.998 & - & \textbf{1.00} \\
\hline
\end{tabularx}
\end{threeparttable}
\label{tab:effective_in_l}
\vspace{-5pt}
\end{table*}

\noindent \textbf{Transductive Settings.} 
We begin by evaluating our verification for Transductive-F. 
Our initial comparison focuses on the \textit{Detection Rate (DR)} under a single verification query to demonstrate the effectiveness of our fingerprinting node selection. 
Table~\ref{tab:effective_trans_f} presents a comparison among our design and two baselines in four datasets against five different attacks. 
The results indicate that our design generally achieves higher performance than the baselines in the majority of cases, while attaining similar performance in the remaining scenarios.
It could also be found that MANC also achieves substantial performance in the Citeseer dataset. 
This outcome is attributed to the fact that this dataset contains approximately 2 to 7 times more node attributes (around 3700) than other datasets (about 500-1400). 
Therefore, simply considering node attributes is sufficient for verification in this context.
However, for graph datasets with fewer attributes, this design approach becomes less effective. 
In practice, graph data may not always possess a large number of node attributes, as they are primarily utilized to represent relationships between items (e.g., the Polblog dataset does not have node attributes). 
Consequently, there is a need for our design, which thoughtfully considers the graph structure when formulating the verification queries.

We then turn to the evaluation of our verification for Transductive-L. 
Since MANC requires access to the internal representation of the model predictions, it is inapplicable when there is limited knowledge of the target model. 
Consequently, we limit our comparison to the \textit{Random Node Selection} method, the results of which are shown in Table~\ref{tab:effective_trans_l}. 
The findings reveal that our design outperforms the baseline designs in the majority of cases. 

\mtrevise{
In addition, we also observe that, Transductive-F tends to have a more stable performance than Transductive-L due to Transductive-F obtaining additional information from the model (i.e., gradient). 
For example, in Table~\ref{tab:effective_trans_f}, the detection rate for Transductive-F in Cora ranges from $0.5-0.96$, while in Table~\ref{tab:effective_trans_l}, they vary between $0.35-1.00$. 
}

\noindent \textbf{Inductive Settings.}
We next evaluate the effectiveness of our design under inductive settings. 
Tables~\ref{tab:effective_in_f} and~\ref{tab:effective_in_l} present the \textit{Detection Rate (DR)} for a single verification query for both the Inductive-F and Inductive-L settings. 
They compare our design with a baseline that queries an ordinary inference graph without modifications, as well as the sensitive sample generation method described in~\cite{HeZL19}. 
It should be noted that in the Inductive-F scenario, where the verifier lacks access to the target GNN models, the sensitive sample generation method cannot be employed. 
Moreover, since this method only takes into account the query samples' attributes (i.e., node attributes), it is also not applicable for graph datasets without node attributes such as Polblog. 

\mtrevise{
The results indicate that, compared to the transductive setting, our design outperforms the baseline method in~\cite{HeZL19}. 
This enhancement in performance is attributed to the unique nature of GNNs, which are more heavily relying on graph structure. 
Thus, constructing an effective fingerprinting graph by modifying both the node attributes and the graph structure is more effective than simply altering the node attributes. 
Furthermore, even when full access to the target model is available, previous designs such as those described in~\cite{HeZL19} may prove unsuitable for GNNs and hence inapplicable. 
In addition, for performance with full access, a minor improvement in the number of queries for GNNs is also significant in real-world deployments, since the nodes issuing queries in GNNs could be physical entities. 
}

\subsection{Efficiency}

\noindent \textbf{Query Efficiency.}
Tables~\ref{tab:efficiency_trans} and~\ref{tab:efficiency_in} present the \textit{QI} for our design, in comparison with two baseline approaches: \textit{Random Node Selection} under the transductive setting, and \textit{Ordinary Inference Graph} under the inductive setting. 
Our design consistently outperforms these baselines across all settings and shows superior performance specifically in inductive settings. 
This corresponds to the effectiveness analysis in Section~\ref{sec:exp_trans}, which illustrates that the construction of a fingerprinting graph by altering the graph structure and node attributes of an ordinary inference graph can be highly effective for verification. 

\begin{figure}[t]
    \centering
    \begin{minipage}[htp]{0.9\linewidth}
        \centering
        \includegraphics[width=0.99\textwidth]{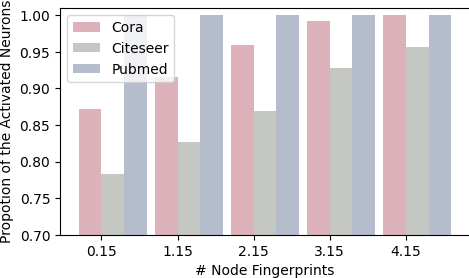}
        

    \end{minipage}
        

    \caption{Activated Neurons of our Node Fingerprints. }
    \label{fig:activated_neurons}
  \vspace{-10pt}
\end{figure}

We also compare our design with the methods detailed in~\cite{HeZL19}, finding that our approach significantly exceeds their performance in most cases, especially when compared to a single query \textit{DR}. 
This is because both designs utilize MANC for the selection of verification nodes, which is based on neuron activation. 
We observe that \textit{DR} using MANC gradually increases with the number of queries. 
This pattern aligns with our rationale described in Section~\ref{sec:transductive_methods}, where the generation of sensitive-samples uses activated neurons to create fingerprints. Figure~\ref{fig:activated_neurons} illustrates activated neurons in GNN models relative to our node fingerprint queries. 
In the context of GNN, the proportion of activated neurons quickly exceeds $85\%$, reducing the effectiveness of selecting sensitive samples according to this criterion. 
Therefore, even when the strategies in~\cite{HeZL19} could achieve a performance similar to our design with a single node query, they will be less efficient when increasing the number of queries to reach higher \textit{DR}. 

\noindent
\mtrevise{
It is also worth noting that even slight improvements over the baseline are significant in practice. 
This is because, in real-world applications, reducing the query number is more vital in GNNs than in DNNs. 
As discussed in Section~\ref{sec:problem_graphaccess}, unlike DNNs where users may issue numerous queries, the nodes initiating queries in GNNs could represent physical entities in a real-world graph (e.g., security appliances in a network graph provided by the system developer to mitigate threats). 
Therefore, minimizing the number of queries becomes essential, accentuating the practical importance of our design in the context of GNNs.
}

\noindent \textbf{Development Cost.}
We then evaluate the time cost of our verification. 
Since our design can be accomplished by issuing inference queries from the fingerprinting nodes in the same manner as inference queries from normal users, the time cost of the queries for our verification is the same as inference queries. 
Namely, our verification does not incur any overhead on the server side. 
%
%
Similarly, on the verifier side, issuing a verification query will also take the same amount of time as a normal prediction query. 
The majority of the time costs come from the generation of the fingerprinting nodes. 
Table~\ref{tab:time cost} shows the time cost of generating one fingerprinting node using both our methods in three datasets. 
All results are recorded when our verification mechanism is running in CPU mode since we assume that the verifier who uses the MLaaS may not have high-performance computation devices such as the GPU. 
Considering that most of our detection requires only a small number of nodes (less than $10$), our fingerprinting node generation methods can be efficient, especially in the transductive setting. 

\begin{table*}[t!]
\centering
\footnotesize
\tabulinesep=1.1mm
\caption{Comparison of Verification Query Number Improvement than Random Node Selection: Our Method vs. MANC for Transductive-F (denoted as F) and Transductive-L (denoted as L).}
\vspace{-8pt}
\begin{threeparttable}
\begin{tabularx}{\textwidth}{c|*{16}{>{\centering\arraybackslash}X}}
\hline
\multirow{3}{*}{} & \multicolumn{4}{c}{Cora} & \multicolumn{4}{c}{Citeseer} & \multicolumn{4}{c}{Polblog} & \multicolumn{4}{c}{Pubmed} \\ \cline{2-17}
 & \multicolumn{2}{c}{MANC} & \multicolumn{2}{c}{Ours} & \multicolumn{2}{c}{MANC} & \multicolumn{2}{c}{Ours} & \multicolumn{2}{c}{MANC} & \multicolumn{2}{c}{Ours} & \multicolumn{2}{c}{MANC} & \multicolumn{2}{c}{Ours} \\ \hline
 & F & L & F & L & F & L & F & L & F & L & F & L & F & L & F & L\\ \hline
BFA & $1.0\times$ & - & $2.7\times$ & $4.0\times$ & $0.8\times$ & - & $1.2\times$ & $1.7\times$ & $2.0\times$ & - & $1.4\times$ & $5.6\times$ & $2.0\times$ & - & $4.0\times$ & $1.5\times$ \\

BFA-F & $1.5\times$ & - & $2.0\times$ & $1.5\times$ & $0.7\times$ & - & $1.0\times$ & $1.1\times$ & $1.5\times$ & - & $1.0\times$ & $2.4\times$ & $1.3\times$ & - & $2.0\times$ & $2.0\times$ \\

BFA-L & $1.4\times$ & - & $1.8\times$ & $7.0\times$ & $0.5\times$ & - & $1.4\times$ & $2.0\times$ & $1.0\times$ & - & $1.0\times$ & $1.0\times$ & $1.0\times$ & - & $3.0\times$ & $1.0\times$ \\

Mettack & $1.0\times$ & - & $2.0\times$ & $1.3\times$ & $2.1\times$ & - & $2.5\times$ & $2.0\times$ & $1.0\times$ & - & $1.2\times$ & $1.2\times$ & $1.2\times$ & - & $1.5\times$ & $1.1\times$ \\

RandAttack & $1.0\times$ & - & $2.0\times$ & $1.3\times$ & $1.2\times$ & - & $4.0\times$ & $3.0\times$ & $1.5\times$ & - & $3.0\times$ & $1.2\times$ & $1.0\times$ & - & $1.0\times$ & $1.1\times$ \\
\hline
\end{tabularx}
\end{threeparttable}
\label{tab:efficiency_trans}
\vspace{-10pt}
\end{table*}

\begin{table*}[t!]
\centering
\footnotesize
\tabulinesep=1.1mm
\caption{Comparison of Verification Query Number Improvement than Random Node Selection: Our Method vs. SS for Inductive-F (denoted as F) and Inductive-L (denoted as L).}
\vspace{-8pt}
\begin{threeparttable}
\begin{tabularx}{\textwidth}{c|*{16}{>{\centering\arraybackslash}X}}
\hline
\multirow{3}{*}{} & \multicolumn{4}{c}{Cora} & \multicolumn{4}{c}{Citeseer} & \multicolumn{4}{c}{Polblog} & \multicolumn{4}{c}{Pubmed} \\ \cline{2-17}
 & \multicolumn{2}{c}{SS} & \multicolumn{2}{c}{Ours} & \multicolumn{2}{c}{SS} & \multicolumn{2}{c}{Ours} & \multicolumn{2}{c}{SS} & \multicolumn{2}{c}{Ours} & \multicolumn{2}{c}{SS} & \multicolumn{2}{c}{Ours} \\ \hline
 & F & L & F & L & F & L & F & L & F & L & F & L & F & L & F & L\\ \hline
BFA & $2.7\times$ & - & $4.2\times$ & $4.1\times$ & $1.2\times$ & - & $3.0\times$ & $3.2\times$ & $3.5\times$ & - & $3.4\times$ & $3.6\times$ & $1.3\times$ & - & $2.6\times$ & $2.6\times$ \\

BFA-F & $1.0\times$ & - & $1.4\times$ & $2.0\times$ & $0.7\times$ & - & $3.2\times$ & $3.0\times$ & $1.5\times$ & - & $1.5\times$ & $1.4\times$ & $1.4\times$ & - & $3.6\times$ & $3.0\times$ \\

BFA-L & $2.3\times$ & - & $2.4\times$ & $2.8\times$ & $0.6\times$ & - & $3.1\times$ & $3.0\times$ & $1.0\times$ & - & $1.0\times$ & $1.0\times$ & $1.1\times$ & - & $2.0\times$ & $2.1\times$ \\

Mettack & $1.1\times$ & - & $2.0\times$ & $1.3\times$ & $1.2\times$ & - & $3.2\times$ & $3.0\times$ & $1.2\times$ & - & $1.3\times$ & $1.7\times$ & $1.4\times$ & - & $2.6\times$ & $2.1\times$ \\

RandAttack & $1.0\times$ & - & $2.0\times$ & $3.3\times$ & $1.4\times$ & - & $2.2\times$ & $2.1\times$ & $2.0\times$ & - & $2.1\times$ & $2.0\times$ & $1.0\times$ & - & $1.0\times$ & $1.0\times$ \\
\hline
\end{tabularx}
\end{threeparttable}
\label{tab:efficiency_in}
\vspace{-10pt}
\end{table*}

\subsection{Dealing with Adaptive Attackers (Robustness)}
\label{sec:exp_theoretic_analysis}

\begin{figure}[t]
    \centering
    \begin{minipage}[htp]{0.44\linewidth}
        \centering
        \includegraphics[width=0.99\textwidth]{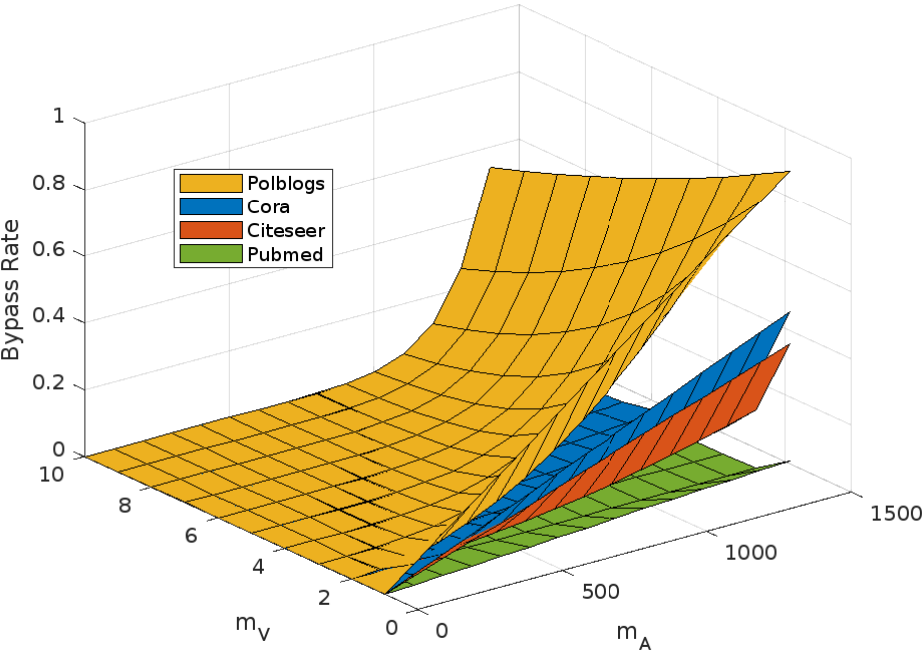}
        
        (a) Transductive Setting

    \end{minipage}
    \begin{minipage}[htp]{0.44\linewidth}
        \centering
        \includegraphics[width=0.99\textwidth]{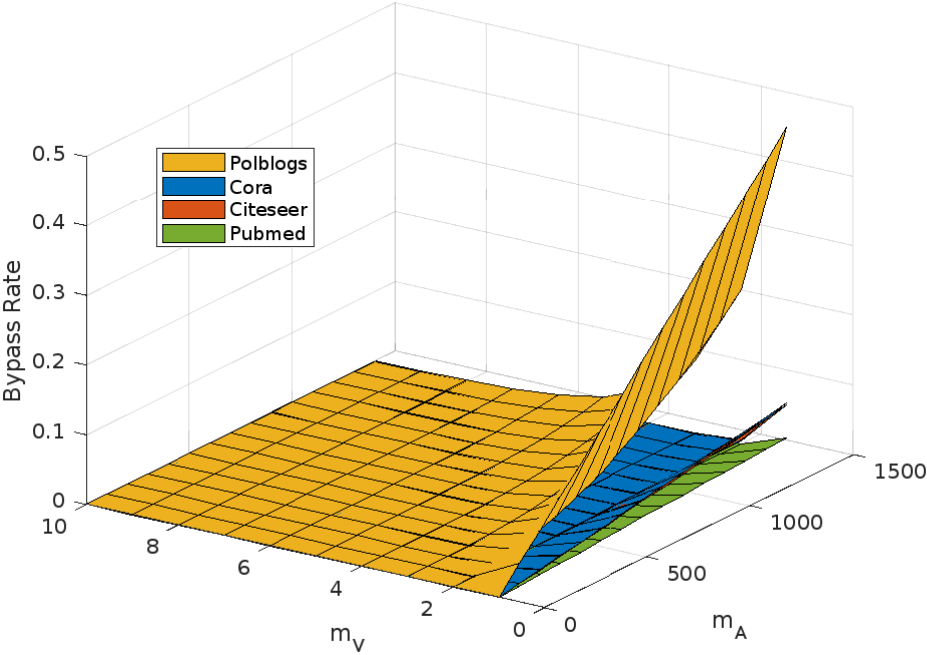}
        
        (b) Inductive Setting

    \end{minipage}
    \caption{Bypassing rates of Adaptive Attackers. }
    \label{fig:Adaptive_analysis}
  \vspace{-10pt}
\end{figure}

In Section~\ref{sec:adaptive_design}, we have introduced how our verification methods in both the transductive and inductive GNN learning settings can deal with adaptive attackers who are aware of the detection and intend to bypass it. 
Here we provide more detailed results about the outcome of our analysis. 

For the transductive setting, we evaluate the bypass rate of an adaptive attacker when he tries to identify the fingerprint nodes. 
Figure~\ref{fig:Adaptive_analysis}(a) show the relationship between the bypass rate and two hyperparameters: $m_A$ (the number of fingerprinting nodes generated by the advanced attacker) and $m_V$ (the number of the fingerprinting nodes generated by the verifier). 
It can be found that, when either $m_V$ increases, the probability that the attacker can identify the fingerprinting node decreases dramatically. 
Meanwhile, if the attacker wants to achieve a high bypass rate, he should increase the number of nodes marked as verification nodes close to the number of total nodes in the graph. 
For example, in the Cora dataset, while honestly responding to more than $2500$ nodes, the attacker can successfully bypass about $20\%$ verification queries. 
Namely, they will honestly respond to more than $90\%$ of the queries which makes their attack much less effective. 
Thus, our randomized node fingerprint can deal with the adaptive attacker. 

Figure~\ref{fig:Adaptive_analysis}(b) shows the relationship between the bypass rate and two hyperparameters in the inductive setting. It can be found that, similar to the results in the transductive setting, the attacker can only achieve a bypass rate of less than $10\%$ by honestly responding to more than $2000$ nodes for all three of our datasets. 
In addition, the verifier can easily reduce the bypass rate by slightly increasing its fingerprinting nodes. 

\subsection{Deployment Cost (Platform-independence)}
\mtrevise{
We finally present the cost of our verification methods. 
As mentioned in Section~\ref{sec:problem}, our verification is designed to be adaptive to different MLaaS platforms. 
%
Firstly, the verifer in our verification process utilizes only query APIs, which are fundamental to MLaaS and supported by all platforms. 
In addition, to deploy our design, all our fingerprint node generation processes are done on the verifier (client) side. 
Namely, our design does not modify or add extra code to the MLaaS server. 
For our verification design for Inductive-F, only around 220 lines of code are deployed on the verifier side, while the method for Inductive-L, only around 170 lines of code are needed. 
}

\begin{table}[t]
\footnotesize
    \label{tab:time cost}
    \centering
    \caption{Time Cost per Fingerprinting Node.}
    \vspace{-8pt}
    \begin{tabular}{c|c|c|c}
    \hline
        Dataset & Knowledge to Models & Transductive  (s) & Inductive  (s) \\
    \hline
        \multirow{2}{*}{Cora} & Full & 3.998 & 23.26 \\
      & Limited & 8.92e-5 & 34.63 \\
    \hline
        \multirow{2}{*}{Citeseer} & Full & 5.148 & 81.24 \\
      & Limited & 7.58e-5 & 34.27 \\
    \hline
        \multirow{2}{*}{Pubmed} & Full & 78.48 & 732.65 \\
      & Limited & 6.06e-5 & 34.26 \\
    \hline
    \end{tabular}
    \vspace{-10pt}
\end{table}

\section{Related Work}
\label{sec:relatedwork}
\label{sec:relatedwork_attacks}
\noindent \textbf{ML Model Verification.}
To address the security threats in MLaaS, recent studies explore how to verify the ML models deployed in the cloud, but none of them is applicable to GNNs. 
Based on the goals of verification, they can be divided into two types, i.e., verifying the model integrity and verifying the model ownership. 
\mainmtrevise{
Prior works propose to utilize traditional software security solutions for ML model integrity verification~\cite{00070L0020,JavaheripiK21}. 
However, they require the verifier to perform the verification which does not align with our assumptions.
%
He \textit{et al.}~\cite{HeZL19} propose a sensitive-sample based method to protect the DNN models in MLaaS. 
Chowdhury \textit{et al.}~\cite{abs-2112-12727} use secret-shared non-interactive proofs to verify integrity.
However, their approach only focuses on DNNs and cannot be extended to GNNs since it neglects to consider the graph structure. 
%
%
Li \textit{et al.}~\cite{00070L0020} propose a lightweight checker-model based verification to verify the integrity of target DNNs. 
However, their proposed validation model needs to be deployed on the server side. 
%
In our work, we assume the cloud is nontrusted and does not require the cloud server for verification. 
}
%
%
Another line of work aims to verify the ownership of an ML model and protect the IP of the ML model. 
Most studies verifying model ownership tend to inject watermarks into models~\cite{0006K21,AdiBCPK18,ZhangGJWSHM18,LiuWZ21,ZhongZ0G020}. 
Note that, they desire to verify the ownership even when the models have been retrained by the attacker, which is different from the breach of the model integrity.

\noindent \textbf{Attacks and Defenses in GNNs.}
Many attacks have been proposed against GNN models in the past few years, including both adversarial attacks and privacy attacks~\cite{abs-2205-07424,XuC0CWHL19,Wu0TDLZ19,ZhangWWYXPY23,WuYPY21}. %
These attacks can be classified into two categories: the attacks perturbing the graph~\cite{XuC0CWHL19,Wu0TDLZ19,WangG19,LiuSZ0H19} and the attacks perturbing the model~\cite{LiuWLX17,RakinHF19,YaoRF20}. 
There are several studies on how to mitigate these attacks on GNNs~\cite{ZhangZ20,ZhuZ0019}. 
Since most of the existing attacks directly perturb the graph data or modify the model parameters by perturbing the training graph, all of these defenses propose to reduce the effects caused by the perturbed graph~\cite{abs-2006-08900,JiangZLTL19,HasanzadehHNDZQ19,abs-2006-08900,Jin0LTWT20}.
We emphasize that these defense methods are different from our design. 
As a general principle, most of the above defense strategies using pre-processing~\cite{Wu0TDLZ19,LuoCYZNCZ21} or reconstructing~\cite{JiangZLTL19,Jin0LTWT20} assume that the defender has full access and control of the target GNNs. 
However, it is not feasible to provide such access in cloud-based GNN services. 
%
As for adversarial training strategies, they either require additional algorithms (such as adversarial regularization terms~\cite{FengHTC21}) or are only suitable for specific attacks~\cite{XuC0CWHL19,ChenLSL20}. 
In addition, such countermeasures are applied before model serving, which cannot directly address threats after the model is deployed in MLaaS. 

Recent studies~\cite{abs-2302-13520,LiuYWYS23} that mitigate BFAs in DNNs are not specifically designed for GNN models. 
These designs often encounter limitations similar to those found in sensitive-fingerprint approaches (i.e., they focus primarily on models with complex structures and overlook the connections among different samples, which are essential in the context of GNNs)~\cite{LiuYWYS23}. Additionally, these methods may require the integration of additional functional components, which makes them less suitable for MLaaS applications~\cite{abs-2302-13520}.


\section{Conclusion}
\label{sec:conclusion}
With the increasing popularity of MLaaS and GNNs, it is essential to explore how to protect the integrity of GNN services against model-centric attacks. 
In this paper, we propose a comprehensive integrity verification framework for GNNs deployed via MLaaS. 
We design a query-based verification mechanism to authenticate the integrity of GNNs through a limited set of fingerprinting nodes. 
Based on two practical (transductive/inductive) settings of node classification tasks and the variant verifier's knowledge, we propose fingerprinting generation algorithms by considering the graph structure and verification nodes' neighbors, respectively.
To address a strong attacker who knows the detection methods and attempts to bypass them, we further enhance our methods by introducing randomized node fingerprints to the verification queries. 
Evaluations of four real-world datasets against five prevalent adversarial attacks demonstrate that our verification method detects them successfully and is $2-4$ times more efficient than the baselines.

\section*{Acknowledgements}
This work is supported in part by a Monash-Data61 Collaborative Research Project (CRP43) and Australian Research Council (ARC) DP240103068,  FT210100097, and DP240101547. Minhui Xue, Xingliang Yuan and Shirui Pan are also supported by CSIRO – National Science Foundation (US) AI Research Collaboration Program. Qi Li's work is in part supported by the National Key Research and Development Project of China under Grant 2021ZD0110502 and NSFC under Grant 62132011. 

\Urlmuskip=0mu plus 1mu\relax
\bibliographystyle{IEEEtran}
\bibliography{reference}

\appendices 

\section{Proposed Attacks}
\subsection{Bit-flip Attack against GNN Models}
\label{app:bfa}
Here we provide the formal definition of BFA. 
Given a two-layer GNN model $f_{\theta}(\cdot)$, the weights are represented as $\theta$. 
An attacker attempts to identify the most vulnerable weight bits, which can maximize the inference loss of GNNs by the perturbed weights.
Such an attack can be formalized as an optimization problem: 
\begin{equation}
    \begin{aligned}
        \max_{\theta'} \quad & \mathcal{L}(f_{{\theta'}}(G),  Y) - \mathcal{L}(f_{{\theta}}(G),  Y) \\
        \mathnormal {s.t. \quad} & B(\theta',\theta)=1,
    \end{aligned}
\end{equation} 
\noindent where $G$ and $Y$ are the graph input and node labels for the inference node sets. 
$\mathcal{L}$ calculates the inference loss between GNN's predictions and target labels. 
$B(\theta',\theta)$ represents the number of flipping bits between the perturbed and clean GNN parameters. 


Overall, we follow the existing strategies of BFA introduced in DNN~\cite{LiuWLX17,BreierHJMB018} and implement them in GNNs. 
Given a set of model parameters, i.e., weights or bias representing a floating number, they are stored as floating-point numbers in the IEEE standard format for floating-point arithmetic. 
%
Considering the 32-bit floating number type, which is a typical number type for the Deep Learning framework, the numbers consist of the sign bit, the exponent bits, and the significant bits. 
To manipulate the value by bit-flipping, we propose to flip the most significant bit of the exponent part. 
\subsection{Poisoning Attacks against GNNs}
\label{app:poisoning_attack_definition}
The attacks that decrease the final model performance by perturbing the graph during the inference period, are called evasion attacks. 
Formally, an evasion attack can be defined as the following optimization problem. 

Given a target graph $G=(\mathcal{V}, \mathcal{E})$, and a well-trained GNN model $f_{\theta}$, an attacker aims at maximising the accuracy loss between the predictions of $f_{\theta}(G)$ and ground truth labels $Y$ by perturbing $G$ to $G'$ within perturbation budget $\epsilon$, as followed: 
\begin{equation}
    \begin{aligned}
        \max_{G'} \quad & \mathcal{L}(f_{{\theta}}(G'),  Y) - \mathcal{L}(f_{{\theta}}(G),  Y) \\
        \mathnormal {s.t. \quad} & D(G', G)<\epsilon,
    \end{aligned}
\end{equation} 
%
where $D(G', G)$ is a perturbation measurement function between two graphs. 

\section{Additional Derivations and Definitions}
\subsection{Fingerprint Score Function for Transductive-F}
\label{app:fingerprinting_proof}
Since the prediction result from $y_i$ is non-linear to the model output $f_{{\theta}'}(G, v_i)$, we approximate the difference between the final prediction labels by the L2 distance between their output confidence vectors, as follows:
\begin{equation}
    \Phi_{i} \propto ||f_{{\theta}'}(G, v_i)-f_{{\theta}}(G, v_i)||_2.
\end{equation}
The perturbation is applied to $\theta$, so we can generate a Taylor expansion of $f_{{\theta}}(G, v_i)$ around current well-trained model parameters ${\theta}$ as follows:
\begin{equation}
\begin{aligned}
& \Phi_{i} \propto \left \|f_{{\theta}+\Delta_{\theta}}(G, v_i)-f_{{\theta}}(G, v_i)\right \|_2 \\
& = \left \|f_{{\theta}}(G, v_i)+{\left(\frac{\partial f_{{\theta}}(G, v_i)}{\partial {\theta}}\right)}^T \Delta_{\theta} + \mathcal{O}({\Delta_{\theta}}^2) - f_{{\theta}}(G, v_i)\right \|_2 \\
& = \left \|{\left(\frac{\partial f_{{\theta}}(G, v_i)}{\partial {\theta}}\right)}^T \Delta_{\theta} + \mathcal{O}({\Delta_{\theta}}^2)\right \|_2, \\
\end{aligned}
\end{equation}
where, $\mathcal{O}(.)$ is the second order (and higher) term of $\Delta_{\theta}$. 
In general, $\Delta_{\theta}$ is a small perturbation as attackers often tend to be in a less noticeable manner. 
Therefore, we disregard this higher-order term and define our fingerprint score:
%
\begin{equation}
    \Phi_{i} = \sum_{\theta_j \in {\theta}}\left \|\frac{\partial \mathcal{L}(f_{{\theta}}(G,v_i),y_i)}{\partial \theta_j}\right \|_2,
    \label{eqt:grad-sensitive_values_app}
\end{equation}
where $\mathcal{L}(.)$ represents the distance between the node prediction and its prediction label $y_i$. 

\subsection{Security Analysis of Randomized Fingerprints}
\label{app:transduct_adaptive}
In this section, we provide a theoretical analysis of the randomized fingerprint method in a transductive setting when facing an adaptive attacker. 
%
%
Here, we define the two processes for verification and adaptive attacks:

\begin{definition}[Verifier $\mathcal{V}$]
Given a graph $G=(\bm{V},\bm{E})$, which has total $N$ nodes ($|\bm{V}|=N$), a verifier uses the fingerprint generation scheme to generate $m_V$ fingerprints among the entire graph. 
The fingerprint node set is noted as $\bm{M}_V$. 
The verifier will use $\bm{M}_V$ to verify the integrity of GNNs. 
\end{definition}

\begin{definition}[Adaptive Attacker $\mathcal{A}$]
An adaptive attacker follows a similar strategy as Verifier $\mathcal{V}$ to generate a set of fingerprints and identify them during the GNN serving. 
Specifically, he first randomly selects $m_A$ nodes as the hypothetical node set. 
Then, he follows the same strategies as Verifier $\mathcal{V}$, to generate $m_A$  hypothetical fingerprints among the entire graph. 
The hypothetical fingerprint set is noted as $\bm{M}_A$. 
The attacker will honestly respond to nodes in $\bm{M}_A$ to bypass the verification. 
\end{definition}

\begin{definition}
[Event $\mathscr{A}$]
Attacker $\mathcal{A}$ successfully identifies all the verification nodes selected by Verifier $\mathcal{V}$ and bypasses the verification, where:
\begin{equation}
    \begin{aligned}
        Pr(\mathscr{A}) = Pr(\bm{M}_V \subseteq \bm{M}_A).
    \end{aligned}
\end{equation}
\end{definition}

Since both $\bm{M}_V$ and $\bm{M}_A$ are randomly sampled from $\bm{V}$, we have: 
\begin{equation}
\label{eqt:p_v2}
    \begin{aligned}
        Pr(\bm{M}_V \subseteq \bm{M}_A) & = \frac{\tbinom{N}{m_A-m_V}}{\tbinom{N}{m_A}} \\
        & = \prod_{i}^{m_V}{\frac{m_A-m_V+i}{N-m_A+i}}
    \end{aligned}
\end{equation}
Thus, when $N \gg m_A > m_V$, we have:
\begin{equation}
\label{eqt:p_v2_final}
    \begin{aligned}
        Pr(\mathscr{A}) & \approx {(\frac{m_A}{N})}^{m_V}
    \end{aligned}
\end{equation}
Such probability becomes negligible when $m_V$ increases unless $m_A \approx N$.
However, if $m_A$ approaches $N$, the attacker will honestly respond to almost all queries, which makes their attack meaningless. 
For example, given a graph with $100$ nodes, if both verifier $\mathcal{V}$ and attacker $\mathcal{A}$ generate their fingerprint set with the size of $5$ ($m_V = m_A = 5$), the probability of $\mathcal{A}$ to bypass $\mathcal{V}$ is about $3.12 \times 10^{-7}$.

\newpage 


\section{Meta-Review}

The following meta-review was prepared by the program committee for the 2024
IEEE Symposium on Security and Privacy (S\&P) as part of the review process as
detailed in the call for papers.

\subsection{Summary}
This paper proposes techniques to mitigate security issues that arise in the use of Graph Neural Networks (GNNs) in Machine Learning as a Service (MLaaS) settings. It uses a query-based verification framework that uses node fingerprint nodes to thwart knowledgeable attackers. Experimental results show that this method efficiently detects several different forms of attack, and outperforms baseline methods by significant margins.

\subsection{Scientific Contributions}
\begin{itemize}
\item Addresses a long-standing issue
\item Provides a valuable step forward in an established field
\end{itemize}

\subsection{Reasons for Acceptance}
\begin{enumerate}
\item This work provides the first systematic investigation study of protecting GNNs in this timely setting against model-centric attacks.
\item The proposed verification framework is comprehensive, covering a variety of access levels and adversarial background knowledge.
\item The paper provides an extensive analysis of the proposed techniques regarding effectiveness and performance.
\end{enumerate}

\balance



\end{document}